\documentclass{aastex631}
\usepackage{amssymb}
\usepackage {romannum}
\usepackage{gensymb}
\usepackage{physics}
\usepackage[autostyle]{csquotes}
\usepackage{soul}
\usepackage[normalem]{ulem}
\usepackage{tikz} 
\usepackage{hyperref}

\begin{document}

\title{ Probing spectral line asymmetries due to the propagating transverse waves in the solar corona}

\author[0000-0002-0786-7307]{Ambika Saxena}
\affiliation{Aryabhatta Research Institute of Observational Sciences (ARIES), Manora Peak, Nainital-263001, Uttarakhand, India}
\affiliation{M. J. P. Rohilkhand University, Bareilly, Uttar Pradesh 243006, India}

\author{Vaibhav Pant}
\affiliation{Aryabhatta Research Institute of Observational Sciences (ARIES), Manora Peak, Nainital-263001, Uttarakhand, India}

\author{Tom Van Doorsselaere}
\affiliation{Centre for mathematical Plasma Astrophysics, Mathematics Department, KU Leuven, Belgium}

\author{M. Saleem Khan}
\affiliation{M. J. P. Rohilkhand University, Bareilly, Uttar Pradesh 243006, India}

\correspondingauthor{Vaibhav Pant}
\email{vaibhav.pant@aries.res.in}







\begin{abstract}

Decades-long studies of asymmetric spectral lines in the solar corona suggest mass and energy transport from lower atmospheric layers to the corona. While slow magnetoacoustic waves and plasma flows are recognized as drivers of these spectral line asymmetries, the role of transverse MHD waves remains largely unexplored. Previous simulations have shown that unidirectionally propagating kink waves, in the presence of perpendicular density inhomogeneities, can produce a turbulence-like phenomenon called ``uniturbulence''.
 However, the spectroscopic signatures of this effect have not been investigated until now. Due to varying Doppler shifts from the plasma elements with different emissions, we expect to observe signatures of both blueward and redward asymmetries. Past instruments like EIS may have missed these signatures due to resolution limitations, but current instruments like DKIST offer a better opportunity for detection. We conducted 3D MHD simulations of transverse waves in a polar plume with density inhomogeneities and performed forward modeling for the Fe XIII emission line at 10749 \AA. Our findings show that transverse waves and uniturbulence induce alternating red and blueward asymmetries, with magnitudes reaching up to 20\% of peak intensity and secondary peak velocities between 30 and 40 km s$^{-1}$, remaining under 100 km s$^{-1}$. These asymmetries propagate with the transverse waves, and even at DKIST resolution, similar signatures can be detected. Our study suggests that spectral line asymmetries can serve as a diagnostic tool for detecting transverse wave-induced uniturbulence.

\end{abstract}



\section{Introduction} \label{sec:intro}
In the late 1970s, the EUV spectrograph on Skylab carried out the first observations of asymmetric spectral line profiles in far-ultraviolet and X-ray emissions from the transition region and the corona of the Sun. These observations showed that the emission line profiles were broadened and contained a secondary broad Gaussian component less intense than the line core \citep{kjeldseth1977emission}.
The High Resolution Telescope and Spectrograph (HRTS) identified these broad components as the signatures of unresolved explosive events \citep{1995SoPh..162..189W}. 

During the last two decades, observations from the Hinode/EUV Imaging Spectrometer (EIS) \citep{article} have indicated the ubiquitous presence of blueward asymmetries in the transition region and coronal lines \citep{2009ApJ...701L...1D, 2009ApJ...707..524M}. These asymmetries are more prominent in loop foot-point regions \citep{2008ApJ...678L..67H, 2010A&A...521A..51P, 2011ApJ...727...58W}, are present for a temperature range of 10,000 K - 2 MK, and can be found in different configurations of magnetic fields, including active regions, coronal holes, and the quiet Sun \citep{2009ApJ...701L...1D,2009ApJ...707..524M}. 
It was suggested that these spectral line asymmetries are due to the overlap of the primary, intense component coming from background emission with a secondary, less intense component accounting for high-speed upflows having speeds in the range of 50 - 150 km s$^{-1}$ \citep{2011tian}. These upflows were spatio-temporally correlated to type II spicules \citep{2007PASJ...59S.655D, 2010A&A...521A..51P} in the chromosphere and may play an essential role in providing mass and energy to the solar corona \citep{2009ApJ...701L...1D, 2009ApJ...707..524M, 2011Sci...331...55D, 2024A&A...689A.135S}. \cite{2011tian} studied the properties of the secondary component using the Modified BR (Blue-Red) analysis technique and found that the Gaussian width of the secondary component is comparable to the primary one. Using simultaneous spectroscopy and imaging, these secondary components were found to be associated with upward propagating disturbances, which were caused by high-speed upflows \citep{2011ApJ...727L..37T, 2009ApJ...707..524M, 2011tian}. 
\cite{2011ApJ...732...84M}, using simulations, demonstrated that strong velocity gradients in the transition region and lower corona contribute to blueward asymmetries at loop footpoints. However, the simulated asymmetries were an order of magnitude less than observations, likely due to the absence of spicules in the model, leaving velocity-gradient-driven upflows as the dominant cause.
\newline
In some studies, slow magneto-acoustic waves were associated with these ubiquitously observed periodic blueward asymmetries \citep{1967ApJ...148..833E, 2010ApJ...724L.194V,1968SoPh....4...96M}. When quasi-static plasma in the observer's LOS is included or when the emission line is averaged over an oscillation period, due to in-phase behavior of velocity and density perturbations, slow waves cause the blue wing of the line to enhance, causing periodic spectral line asymmetries \citep{2010ApJ...724L.194V}. 
\cite{2012ApJ...759..144T}, using Hinode/EIS observations, reported no detectable signatures of spectral line asymmetries by transverse MHD waves in closed coronal loops.

High resolution observations of several chromospheric and coronal structures have emphasized the presence of inhomogeneities perpendicular to the magnetic field direction (see the review by \citet{2023RvMPP...7...17M}).
Recent high-resolution EUV observations have revealed that polar coronal plumes are made of small thread-like filamentary structures that act as waveguides for MHD waves \citep{2009Gabriel,2021ApJ...907....1U,2023ApJ...954...90M}. Simulation-based studies \citep{2017NatSR...714820M,magyar2019understanding,2016ApJ...823...82M}, which are based on and have utilized perpendicular inhomogeneities in their models, have highlighted the formation of turbulence due to unidirectional propagation of transverse MHD waves in the presence of inhomogeneities perpendicular to the magnetic field. These inhomogeneities lead to generalized phase mixing, which causes a turbulence-like behavior called \enquote{Uniturbulence}. \cite{Pant2019InvestigatingE} extended this model to include gravity and studied the effects of uniturbulence in generating the observed spectroscopic features. Using the same model, the variations in the filling factors of overdense plasma structures in the solar corona by the turbulence driven by MHD waves were investigated \citep{Sen_2021}.
Due to the limited capabilities of current instruments, the signatures of uniturbulence cannot be deduced from imaging and conventional spectroscopic methods. \cite{Pant2019InvestigatingE} found that line-of-sight superposition of plasma emission together with uniturbulence can lead to spectral line broadening. However, broadening alone is not conclusive evidence of uniturbulence. In contrast, spectral line asymmetries may better reflect fine-scale plasma dynamics and, if present, could serve as potential indicators of Alfvén wave–driven turbulence. Furthermore, spectral line asymmetries caused by propagating transverse waves in polar plumes have not yet been reported.
Our study investigates the spectral line asymmetries caused by propagating transverse waves, using MHD simulations in a coronal polar plume having density inhomogeneities perpendicular to the magnetic field. In this work, we have used the model developed by \cite{Pant2019InvestigatingE} and performed forward modeling to study the spectroscopic signatures of propagating transverse waves and developed uniturbulence. The paper is organized as follows: Section \ref{sec:style} describes the simulation setup and forward modeling of the data. Section \ref{sec:Data Analysis and Results}, briefly describes the BR asymmetry technique and data analysis, and Section \ref{sec:cite} discusses our results. 

\section{Simulated and Forward Modeled Data} \label{sec:style}

\begin{figure*}[ht!]
    \centering
    \includegraphics[angle=0,scale=0.5]{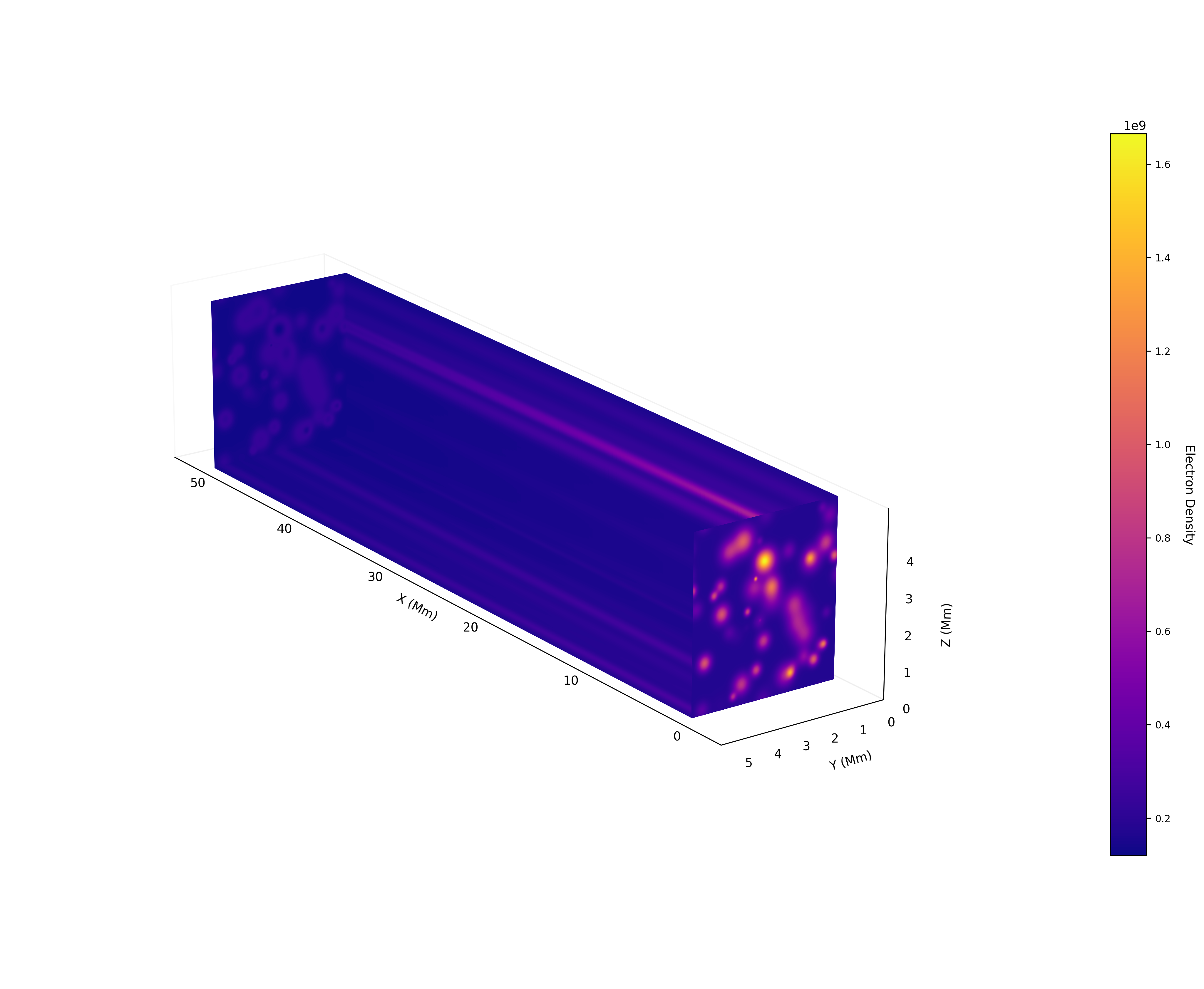}
   
    \caption{Full simulation cube at $t = 0$. The $x-$ axis represents the radial direction. The density stratification is evident along the radial direction, while density inhomogeneities can be seen in the $y-z$ plane \citep{Pant2019InvestigatingE}. }
    \label{fig:simulatio cube}
\end{figure*}

We have used the data from the simulations performed by \cite{Pant2019InvestigatingE} and are discussing the simulation setup for the sake of completeness. The ideal 3D MHD simulations were performed in the Cartesian geometry using MPI-AMRVAC \citep{2014ApJS..214....4P} for the transverse MHD waves in the coronal polar plumes. The grid was uniform without any mesh refinement having size $128\times 512 \times 512 $ spanning 50 Mm$\times$ 5 Mm $\times$ 5 Mm, see Figure \ref{fig:simulatio cube}. The following MHD equations were solved for the simulation:
\begin{equation}
 \frac{\partial \rho}{\partial t}  + \div({\rho \textbf{v}})=0  ,
\end{equation}

\begin{equation}
 \frac{\partial (\rho \textbf{v})}{\partial t}  + \div({\rho \textbf{v}\textbf{v}- \textbf{B}\textbf{B}/{\mu_0}}) + \grad({p+ \textbf{B}^2/{2\mu_0}}) -\rho \textbf{g} =0  ,
\end{equation}

\begin{equation}
 \frac{\partial E}{\partial t}  + \div[({\textbf{v} E- \textbf{B}\textbf{B}\cdot \textbf{v}/{\mu_0}}) + \textbf{v}\cdot(p+ \textbf{B}^2/{2\mu_0})]-\rho \textbf{v}\cdot \textbf{g}  =0  ,
\end{equation}

\begin{equation}
 \frac{\partial \textbf{B}}{\partial t}  - \curl({\textbf{v}\times \textbf{B}})=0  ,
\end{equation}

\begin{equation}
 \div{\textbf{B}}=0  ,
\end{equation}

Here $\rho$ is the mass density, $\textbf{B}$ is the magnetic field, $p$ is the gas pressure, and \textbf{$E$} is the total energy density. The acceleration due to gravity $\textbf{g}$ acts along the negative $x$-axis of the simulation setup, which is the radial direction of the solar atmosphere. Here, gas pressure follows the ideal gas law given as $p=\frac{k_b}{\mu m_H}\rho T $ and total energy density is given by $E=\frac{p}{\gamma-1}+\frac{\rho \textbf{v}^2}{2}+\frac{\textbf{B}^2}{2\mu_0}$. Here, \textbf{$u_0$} is the magnetic permeability of the free space, $m_H$ is the proton's mass, and $k_b$ is the Boltzmann constant. The value of the adiabatic constant $\gamma$ is considered $5/3$, while the value for coronal abundance, $\mu$, is chosen to be 0.6.

The simulations were set up for a vertically isothermal atmosphere, causing gravitational stratification, leading to the exponential decay of density. A magnetic field of magnitude $\approx$ 5 G is present, which is initially uniform and vertical. Initially, inhomogeneities in density are randomly introduced perpendicular to the magnetic field according to the following expression \citep{Pant2019InvestigatingE} :
\begin{equation}
    \rho(x,y,z)=\\ \left(\rho_0 +\sum_{i=0}^{50} A_i \exp^{-[(y-y_i)^2+(z-z_i)^2]/{2\sigma_{i}^{2}}} \right )\exp^{-{x/H(y,z)}}  .  
\end{equation}
Here, $\rho_0$ is the background density with value $2 \times 10^{-13}$  kg m$^{-3}$, $H(y, z)$ is the temperature-dependent scale height, which has different values at different locations in the $y-z$ plane. $A_i$ denotes the magnitude of inhomogeneity, which is drawn randomly from a uniform distribution of $[0,5] 
 \rho_0$. $\sigma_i$ denotes the spatial extent of the inhomogeneity, randomly drawn from a uniform distribution of $[0, 250]$ km.
The spatial location of inhomogeneity is randomly chosen from a uniform distribution of $[-2.5, 2.5]$ Mm within the simulation domain. The plasma $\beta$ is assumed to be low, $\beta=0.15$. The simulations were evolved for $\approx$ 100 s to relax the initial state to pressure equilibrium before implementing driving. All the boundaries are kept open at this stage so that any MHD wave generated can leave the simulation domain. Here, $t = 0$ denotes the time when simulations reach the pressure equilibrium. After this step, the entire bottom boundary ($x =$ 0 Mm) of the cube was driven to excite transverse MHD waves using velocity drivers of the following form :
\begin{equation}
    v_y(x=0,t) = \sum_{i=1}^{10} U_i \sin(\omega_{i} t),
\label{eq:eq7}
\end{equation}

\begin{equation}
    v_z(x=0,t) = \sum_{i=1}^{10} V_i \sin(\omega_{i} t).
\label{eq:eq8}
\end{equation}

The $y$ and $z$ boundaries at this state are periodic, while the top boundary is kept open, so these waves can leave the domain. $U_i$ and $V_i$ are chosen from a uniform distribution $[-U_0,U_0]$. We are primarily using data for $U_0: 11/\sqrt{2}$ km s$^{-1}$ and for some 
 specific cases we have used 
 the data for  $U_0: 22/\sqrt{2}$ km s$^{-1}$. In Equation \ref{eq:eq7} and \ref{eq:eq8}, the wave periods are considered from log normal distributions, centered at a mean value of $400\,\mathrm{s}$ obtained by statistical studies of transverse motions of polar plumes \citep{2014ApJ...790L...2T,2015NatCo...6.7813M}. The rms value of the velocity driver (vrms) is calculated by averaging the driver's velocity over the complete  
bottom boundary, which corresponds to $\sim$ 15 km s$^{-1}$ for $U_0: 11/\sqrt{2}$ km s$^{-1}$ and 26 km s$^{-1}$ for $U_0: 22/\sqrt{2}$ km s$^{-1}$, respectively. The initial state at $t$=0 and evolved state at $t$= 45 of the cross section at $x$= 20 Mm in terms of density and temperature structures can be seen in figure 4 of \citet{Pant2019InvestigatingE}. In the model, the average loop temperature is approximately $2.8\,\mathrm{MK}$. The temperatures within the inhomogeneous regions range from $0.5$ to $2\,\mathrm{MK}$, whereas the background plasma maintains higher values in the interval $3.5$–$4\,\mathrm{MK}$.  

There are 50 snapshots for every run with a cadence of 20s. Hence, the total duration of simulations is 1000s. At $t = 0$, the average sound speed and the Alfv\'en speeds are $\approx$ 120 km s$^{-1}$ and 500 km s$^{-1}$, respectively.    
The FoMo tool \citep{2013A&A...555A..74A, 2016FrASS...3....4V} was used to forward model the optically thin emission from the coronal plasma, which was assumed to be at ionization equilibrium. The emission was calculated for every grid point using:
\begin{equation}
    \epsilon(x,y,z,t)=  \frac{A_b}{4\pi}  n_e^2(x,y,z,t)G(n_e,T)  ,
\end{equation}
where $A_b$ is the coronal abundance, $n_e$ is the electron density, and $G(n_e, T)$ is the contribution function for Fe XIII, is computed using the CHIANTI atomic database \citep{1997A&AS..125..149D}.
The physical parameters were converted to spectroscopic variables for the Fe \Romannum {13} emission line centered at 10749 {\AA}. The main motivation for selecting this spectral line is our interest in determining whether the signatures of these asymmetries would be detectable at the resolution of DKIST instrument. The formation temperature of the line is $\approx 1.6 $ MK and the thermal line width is $\approx$ 21.7 km $s^{-1}$. The emission was synthesized by integrating the specific intensity along multiple lines of sight (LOS). We considered 12 LOS orientations in the $y$--$z$ plane, corresponding to the cross sectional plane of the loop, with viewing angles of $0^{\circ}$, $15^{\circ}$, $30^{\circ}$, $45^{\circ}$, $60^{\circ}$, $75^{\circ}$, $90^{\circ}$, $105^{\circ}$, $120^{\circ}$, $135^{\circ}$, $150^{\circ}$, and $165^{\circ}$. These are hereafter referred to as the ``angles across the magnetic field direction.'' Additionally, 9 LOS orientations were selected along the magnetic field direction ($x$-axis), with viewing angles of $45^{\circ}$, $50^{\circ}$, $55^{\circ}$, $60^{\circ}$, $65^{\circ}$, $70^{\circ}$, $75^{\circ}$, $80^{\circ}$, and $85^{\circ}$, which we refer to as the ``angles along the magnetic field direction.'' LOS integration reduced the data cube to artificial images with two spatial dimensions. Every point in this integrated data corresponds to an artificial spectrum (noiseless and free of any background coronal emission) that has a spectral resolution of $0.146$ \AA, i.e, a velocity resolution of 4.088 km s$^{-1}$. The spectra contain 50 points within the wavelength range $10745.4$\,\AA\ to $10752.6$\,\AA.

\section{Data Analysis and Results} \label{sec:Data Analysis and Results}

\subsection{The BR Asymmetry Technique} \label{sec:br}

The original BR asymmetry technique was introduced by \cite{2009ApJ...701L...1D}.
In this technique, the emission from the blue wing of the spectra is subtracted from the red wing within the same wavelength (velocity) range at a fixed wavelength (velocity) from the centroid. From this line centroid, the range of integration is successively stepped outward, and a BR asymmetry profile is formed. This technique was later modified by \cite{2011tian}, in which the center of the spectra is chosen to be the peak or maximum of the spectra rather than the centroid of the fitted Gaussian over the spectra.

The BR asymmetry is quantitatively calculated by the expression given by \cite{2009ApJ...701L...1D}:

\begin{equation} 
    \label{eq3}
    BR_{v_s}= \int_{v_s - \delta{v/2}}^{v_s +\delta{v/2}} I_{v_s} \,d_{v} \ -\int_{-v_s - \delta{v/2}}^{-v_s +\delta{v/2}} I_{v_s} \,d_{v} \  
    \end{equation}\ \  
, where $v_s$ is the velocity from the line centroid, $\delta{v}$ is the velocity range over which BR Asymmetry is determined, $I_{v_s}$ is the spectral intensity. We used $v_s =$ 1 km s$^{-1}$ and $\delta{v}=$ 20 km s$^{-1}$ for our analysis. We have performed single Gaussian fits over each spectrum obtained after the LOS integration.
The primary line parameters, such as intensity, line width, and Doppler shifts, were obtained from this fitting.
 We used the Modified BR asymmetry technique \citep{2011tian} to quantitatively analyze the spectral line asymmetries. For each spectra, for every step in $v_s$, a BR asymmetry value is obtained using Equation (\ref{eq3}). In the subsequent analysis, we form a velocity vs BR asymmetry profile, which illustrates the variation of blue-to-red wing intensity with the velocity from the peak of the spectra, denoted by $v_{s}$. The BR asymmetry values are normalized to the peak intensity and are denoted by  $\frac{B-R}{I}$.
In this study, we analyzed the data for a single line of sight for the case of two different velocity drivers having the rms velocities of $15$ km s$^{-1}$ and $26$ km s$^{-1}$. We also studied the case for a combination of various randomly chosen lines of sight to replicate the observations of the optically thin corona. We interpolated the spectra to 10 times their original resolution before performing the BR asymmetry analysis. 

\subsection{Spectral line asymmetry for a single LOS }

\begin{figure*}[ht!]
    \centering

    \begin{minipage}{0.48\textwidth}
        \centering
        \begin{tikzpicture}
            \node[inner sep=0pt] (img1) {\includegraphics[scale=0.9]{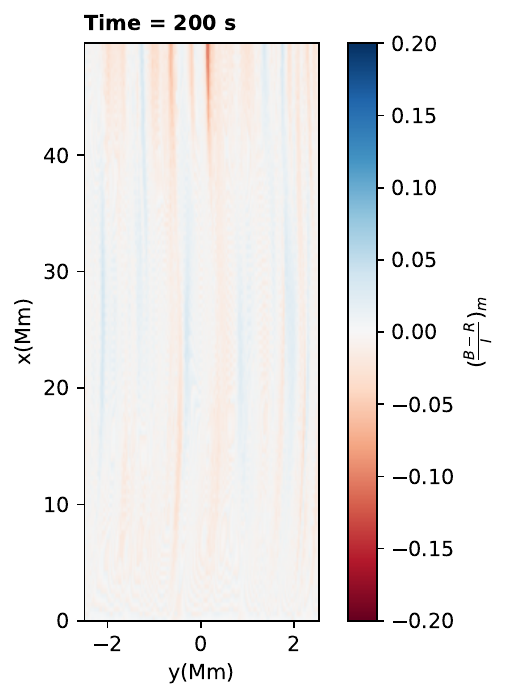}}; 
            \node[anchor=north west] at (img1.north west) {\textbf{(a)}};
        \end{tikzpicture}
    \end{minipage}%
    \hfill
    \begin{minipage}{0.48\textwidth}
        \centering
        \begin{tikzpicture}
            \node[inner sep=0pt] (img2) {\includegraphics[scale=0.9]{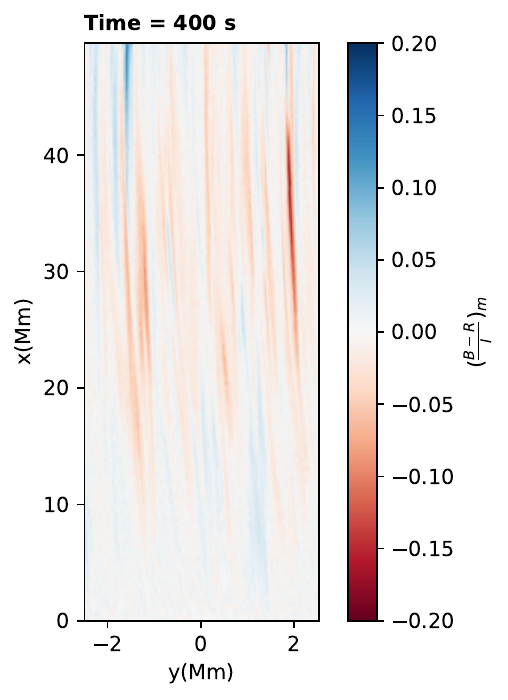}}; 
            \node[anchor=north west] at (img2.north west) {\textbf{(b)}};
        \end{tikzpicture}
    \end{minipage}

    \vspace{0.3cm} 

    \begin{minipage}{0.48\textwidth}
        \centering
        \begin{tikzpicture}
            \node[inner sep=0pt] (img3) {\includegraphics[scale=0.9]{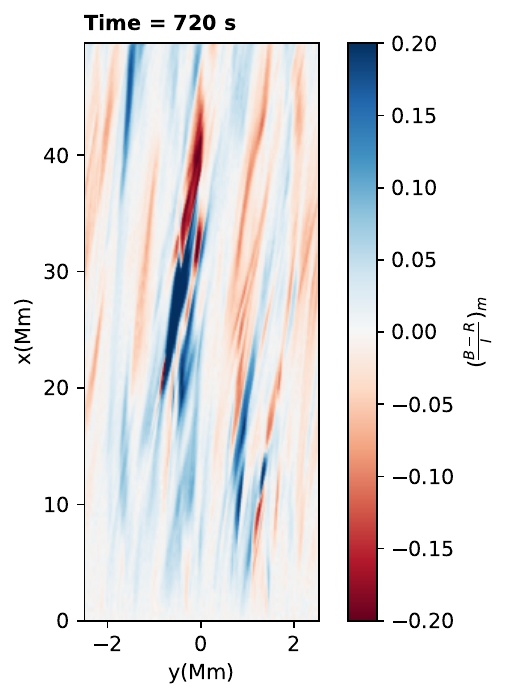}}; 
            \node[anchor=north west] at (img3.north west) {\textbf{(c)}};
        \end{tikzpicture}
    \end{minipage}%
    \hfill
    \begin{minipage}{0.48\textwidth}
        \centering
        \begin{tikzpicture}
            \node[inner sep=0pt] (img4) {\includegraphics[scale=0.9]{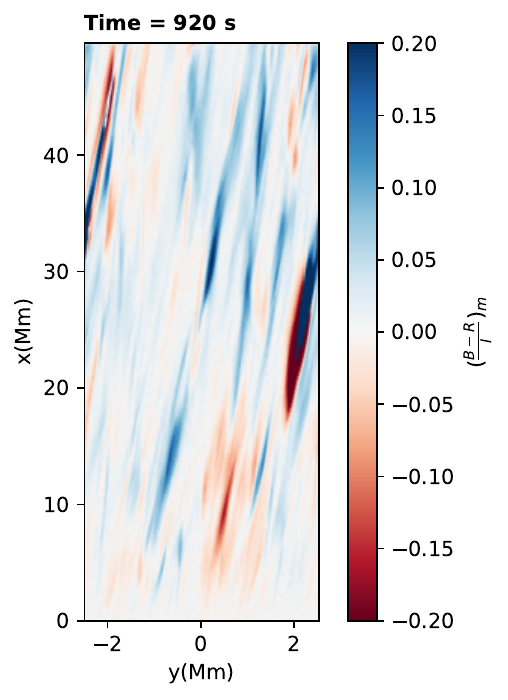}}; 
            \node[anchor=north west] at (img4.north west) {\textbf{(d)}};
        \end{tikzpicture}
    \end{minipage}

    \caption{Spatial maps of BR asymmetry, ${(\frac{B-R}{I})_m}$ values,
    for LOS integrated emission along $z$-axis, for a) $t$ = 200 s, b) 400 s, c) 720 s, and d) 920 s of the simulation, where the velocity driver has speed, $v_{rms}$ = 15 km s$^{-1}$.}
    \label{fig:brmaps}
\end{figure*}
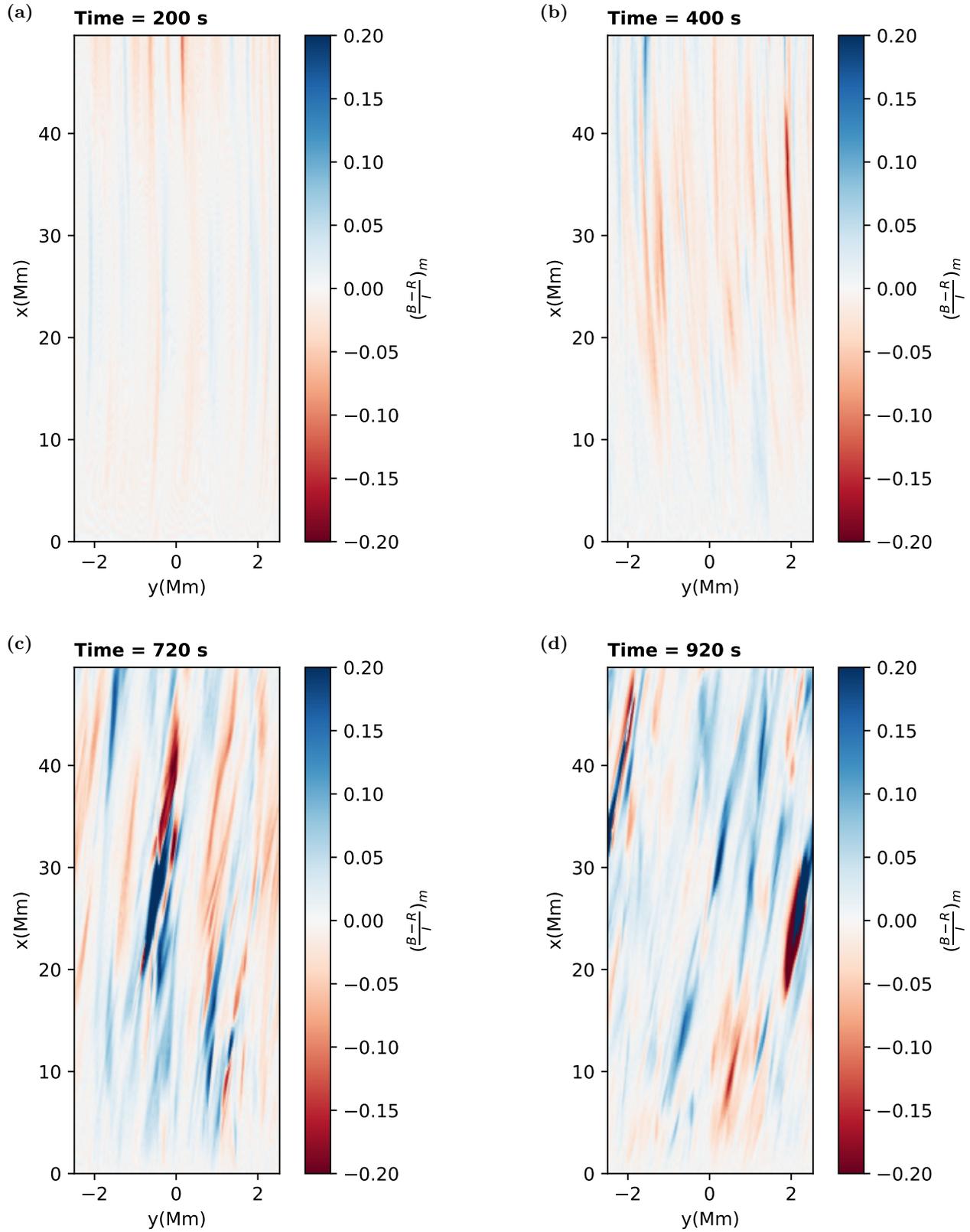

\begin{figure*}[ht!]
    \centering
    \makebox[\textwidth][c]{%
      \begin{minipage}{0.46\textwidth}
        \centering
        \begin{tikzpicture}
          \node[inner sep=0pt] (img1) {\includegraphics[width=\linewidth]{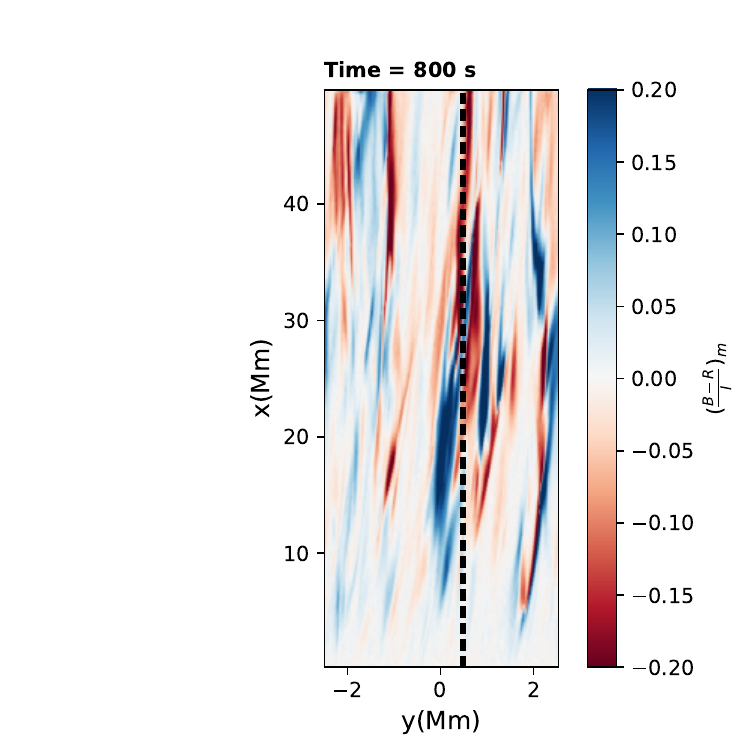}}; 
          \node[anchor=north west] at (img1.north west) {\textbf{(a)}};
        \end{tikzpicture}
      \end{minipage}\hspace{0.03\textwidth}%
      \begin{minipage}{0.46\textwidth}
        \centering
        \begin{tikzpicture}
          \node[inner sep=0pt] (img2) {\includegraphics[width=\linewidth]{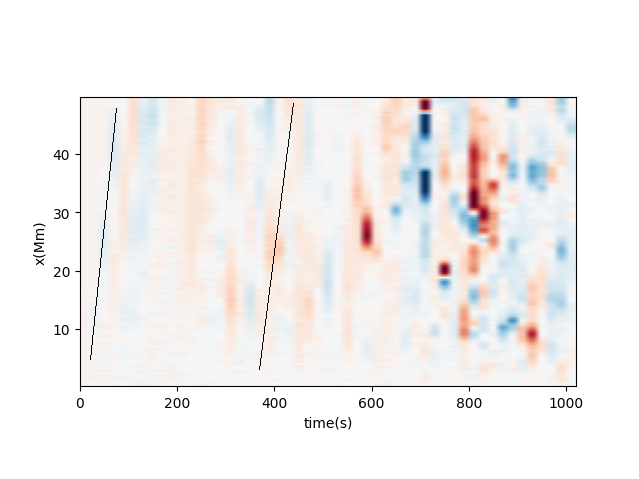}}; 
          \node[anchor=north west] at (img2.north west) {\textbf{(b)}};
        \end{tikzpicture}
      \end{minipage}%
    }

    \caption{Distance--time maps for the BR asymmetry features for $v_{\mathrm{rms}} = 15~\mathrm{km~s^{-1}}$: 
    (a) The slit position is depicted by the dashed line on the BR asymmetry map, and 
    (b) the distance--time map of the features along the slit length shown in (a). 
    The black solid lines depict the slopes of the propagating BR asymmetry features.}
    \label{fig:brxt}
\end{figure*}

We first analyzed the data for the optically thin emission integrated along a single line of sight ($LOS = 0\degree$), i.e., along the $z$-axis. We obtained the BR asymmetry maps using the signed maximum/minimum values of the BR asymmetry obtained for different velocity ranges using Equation (\ref{eq3}) for each spectrum, which we will denote by $(\frac{B-R}{I})_m$, throughout the text.  We studied the variation of these spectral line asymmetries with height and time. We have further obtained the statistical properties of these transverse wave-induced spectral line asymmetries. 
Figure \ref{fig:brmaps} shows the maps of BR asymmetry for different time instances of the simulation. The presence of both blue and red spectral line asymmetries can be seen, which can be observed as alternating red and blue ridges in the distance-time ($x-t$) map and strong asymmetry patches start to appear at later times, which could be seen in Figure \ref{fig:brxt}, panel (b), from around $t$ = 600-1020 s. These patches also seem to propagate upward along the $x$ direction, which can be seen in the $x-t$ map in Figure \ref{fig:brxt} . These $x-t$ maps are created by keeping a 5-pixel wide virtual slit at {$y=$ 0.45 Mm,} on a BR asymmetry map for $t$=800s; here, the dashed line represents the location of the artificial slit.

To understand the time evolution of these asymmetries, we analyzed the emission from the transverse cross-section of the cube at a fixed height. These emission maps were over-plotted with the magnitude and the direction of the velocity of the displaced plasma (blue arrows), a reference arrow is provided at the top of the emission maps. These maps revealed that the asymmetries are caused by overlapping inhomogeneous structures with varying emissions along the LOS that move with velocities different from the background plasma.

\subsubsection{Variation of the spectral line asymmetry with height and time}
\begin{figure}[!ht]
    \centering
    \begin{interactive}{animation}{figure4_5.mp4}
    \includegraphics[width=\textwidth]{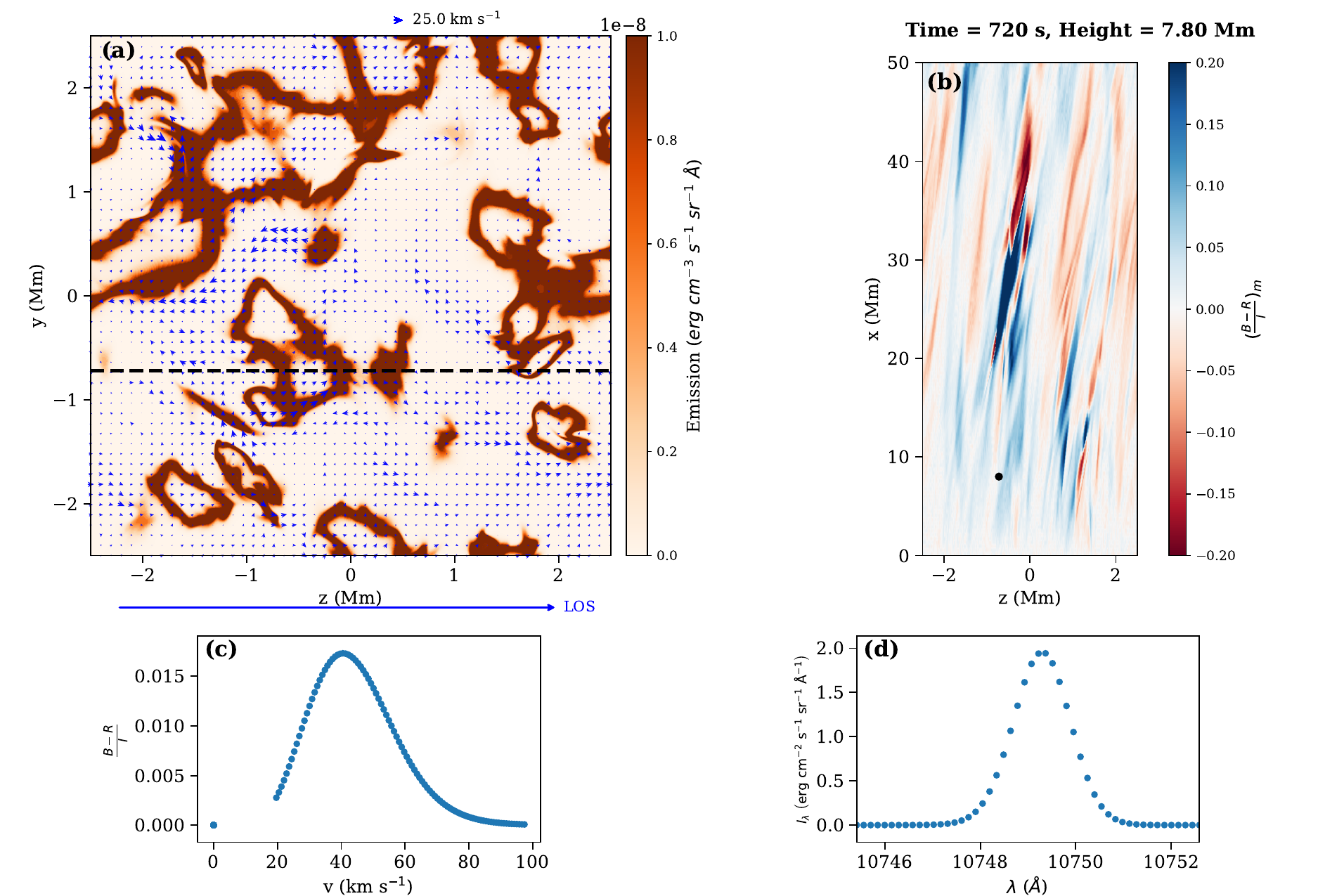}
    \end{interactive}
    \caption{
    Variation of BR asymmetry with height for LOS integrated emission along the $z$-axis, across the guided magnetic field direction, $v_{rms} = 15$ km s$^{-1}$, $t = 720$ s, and $x = 8.01$ Mm:  
    (a) Emission profile of the transverse cross-section of the simulation cube for the Fe XIII emission line, over-plotted with blue arrows representing plasma velocity (the large horizontal arrow indicates the LOS along the $z$-axis; dashed line marks $y = -0.72$ Mm), a reference arrow to compare speeds is provided at the top of the emission map.  (b) Spatial 2D map of ${(\frac{B-R}{I})_m}$ for the full simulation domain, with the black dot showing the location of the spectral line asymmetry derived from LOS integration along the dashed line in (a).  (c) Velocity vs. normalized BR asymmetry, $\frac{B-R}{I}$, profile at the black dot location in (b), where $v_s$ is the velocity from the spectral peak.  (d) Spectrum at the same location as in (b), with $I_{\lambda}$ representing specific intensity.  
    A full animation of BR asymmetry evolution with height is provided in the online version of this article.
    }
    \label{fig:brh20}
\end{figure}

We first studied the variation of spectral line asymmetries with the height, $x$ in the solar atmosphere. In Figure \ref{fig:brh20}, on the top left panel, the direction and magnitude of the plasma velocity (blue arrows) are plotted over the emission in the Fe XIII line for the cross-section of the cube at height, $x$ = 8.01 Mm and $t$= 720 s. 
The vertical dashed black line represents the location $y$ = -0.72 Mm at which the emission is integrated for the LOS along the $z$- axis. The  BR asymmetry map is plotted on the top right panel for the same instant and complete spatial domain of the simulation. The black dot on the BR asymmetry map is at the location $y$ = -0.72 Mm and $x$ = 8.01 Mm, depicting the maximum spectral line asymmetry, where the emission from the dashed line in the Fe XIII emission maps on the left panel of Figure \ref{fig:brh20} is integrated.
 The value of ${(\frac{B-R}{I})_m}$, is 0.017 at the location of the black dot. This can be seen from Figure \ref{fig:brh20}, panel (c), where the normalized BR asymmetry, $\frac{B-R}{I}$, vs velocity, is plotted at the location of the black dot, which represents value of ${(\frac{B-R}{I})_m}$ in the BR vs velocity map. The observed data gap in this plot is due to the chosen strip width, $\delta v$, used in calculating the BR asymmetry. With a strip width of $20$~km~s$^{-1}$, the calculation begins at the line center and progresses 
toward both wings. The first value corresponds to the velocity at the center of the strip 
($10$~km~s$^{-1}$), after which the strip is shifted in increments of $1$~km~s$^{-1}$. 

\begin{figure*}[ht!]
    \centering
    \includegraphics[width=\textwidth, keepaspectratio]{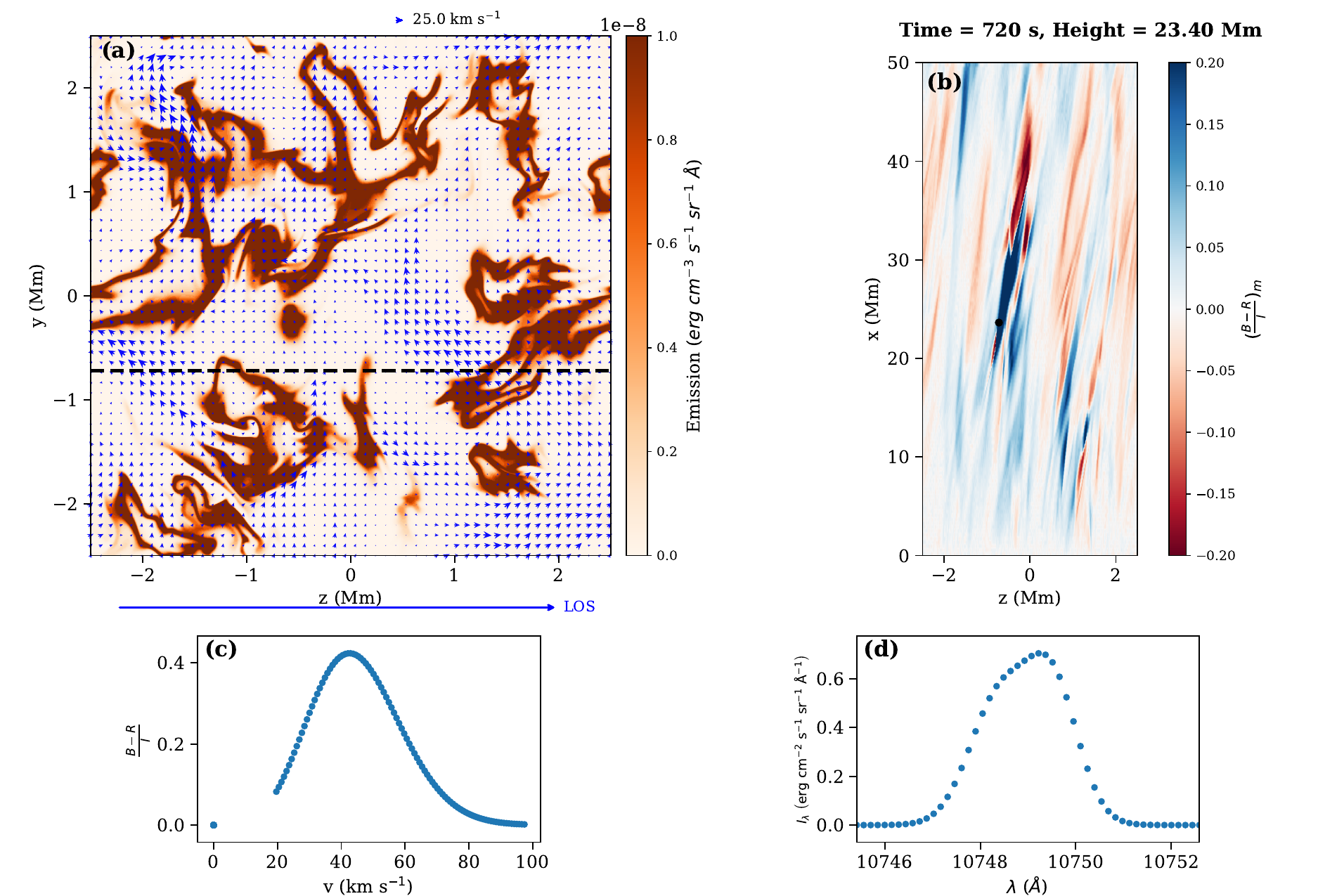} 
    \caption{Same as Figure \ref{fig:brh20} but for $x$ = 23.63 Mm.}
    \label{fig:brh57}
\end{figure*}

We have also plotted the respective spectrum for the same location, which looks slightly asymmetric towards the blue side; see Figure \ref{fig:brh20} bottom right panel.
In Figure \ref{fig:brh57}, we have performed a similar analysis at $x$ = 23.63 Mm, 
at the same $y$ location and time. Here, in panel (a), we can see that the plasma density has decreased due to the increase in height; however, the developed turbulence is more prominent than in the earlier case, primarily because of the increase in wave amplitude caused by the decrease in density. This led to the formation of more small-scale features. Here, the value of ${(\frac{B-R}{I})_m}$ is 0.424, visible from the $v_s$ vs $\frac{B-R}{I}$ plot in Figure \ref{fig:brh57}, panel(b).
In the emission plots in Figure \ref{fig:brh20}, at $y$ = -0.72 Mm, the boundaries of inhomogeneities, which have spread out in size due to the development of turbulence, are comparatively blue-shifted along the LOS, compared to the background plasma. This gives rise to a blueward asymmetry in the spectra in these regions. In Figure \ref{fig:brh57}, we can see a similar but more prominent effect due to the further development of the turbulence, thereby giving rise to a more substantial blueward asymmetry. A strong patch of blue asymmetry is present in the spatial BR asymmetry map in panel (b), at the location of the black dot; above the blue patch is a slightly less intense red patch. In this case, we can note both blueward and redward alternating asymmetry as we move along the height at a fixed location along the $y$ - axis.

\begin{figure*}[ht!]
    \centering
    \begin{interactive}{animation}{figure6_7.mp4}
     \includegraphics[width=\textwidth, keepaspectratio]{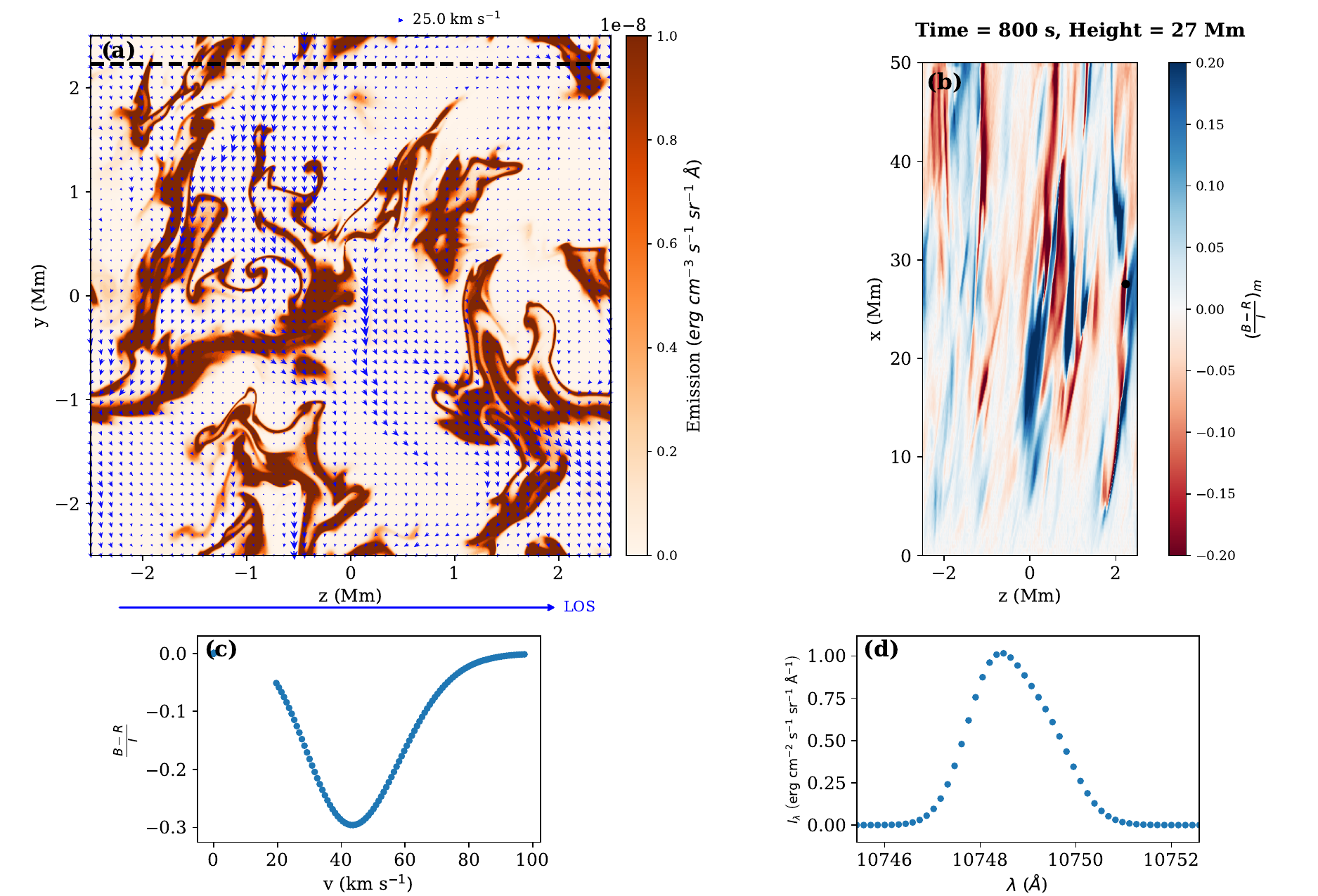} 
        
    \end{interactive}
   
    \caption{Same as Figure \ref{fig:brh20}, but for $x$ = 27.54 Mm, $z$ = 2.23 Mm, $v_{rms}$ = 15 km s$^{-1}$ and $t$ = 800 s. 
    A full animation for the time evolution of BR asymmetry is provided in the online version of this article.}
    \label{fig:brt40}
\end{figure*}

\begin{figure*}[ht!]
    \centering
    \includegraphics[width=\textwidth, keepaspectratio]{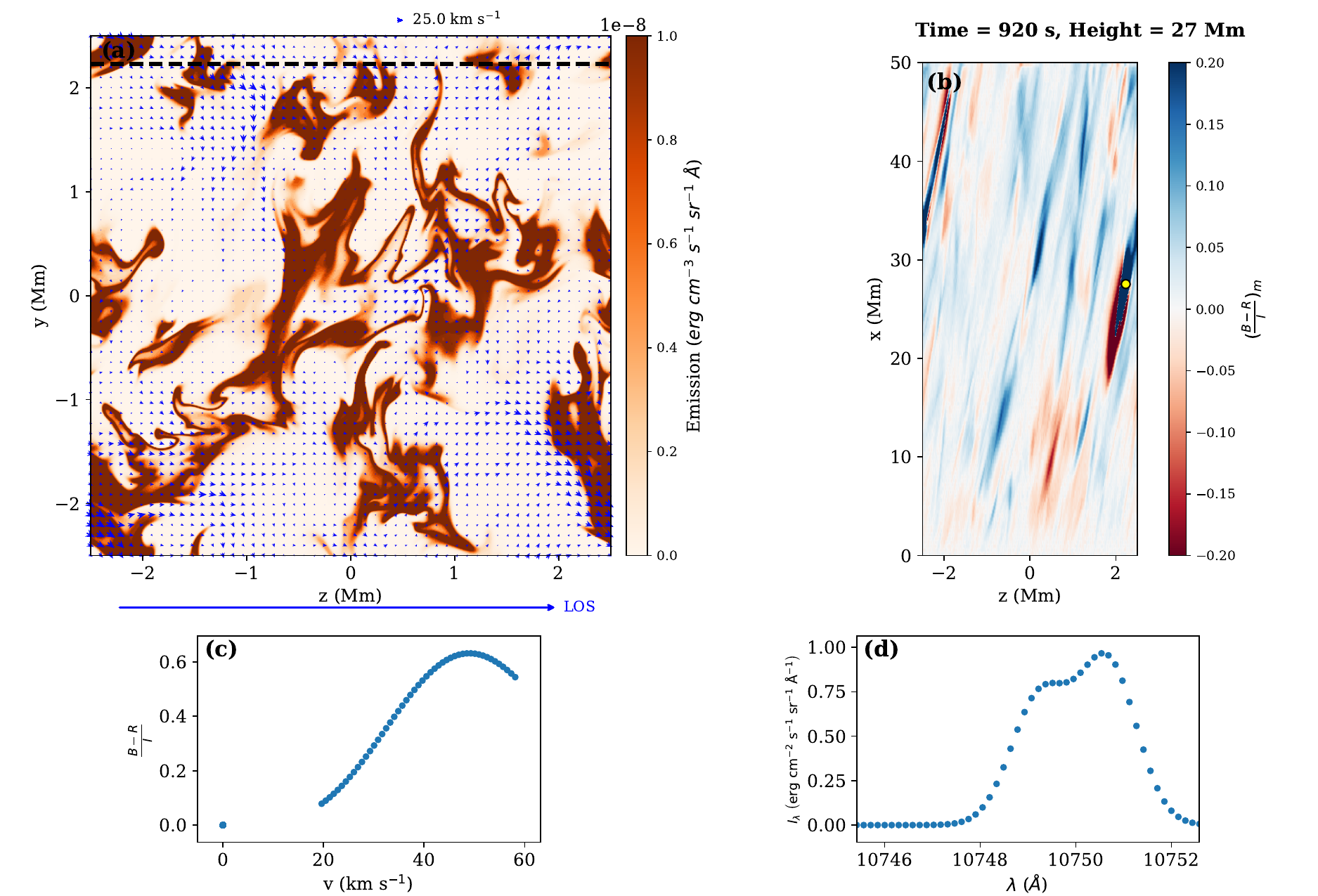} 
    \caption{Same as Figure \ref{fig:brh20} but for $x$ = 27.539 Mm, $z$ = 2.23 Mm, $v_{rms}$ = 15 km s$^{-1}$, and $t$ = 920 s.}
    \label{fig:brt46}
\end{figure*}

We performed the same analysis to study the variation of spectral line asymmetries with time. In Figure \ref{fig:brt40}, the Fe XIII emission map at panel (a), at $y$ = 2.23 Mm and $t$ = 800s, shows a spatial distribution of inhomogeneities and a strong emission from the inhomogeneity boundaries can be seen. For more details regarding emission maps, see Section 4 of \citet{Pant2019InvestigatingE}.  
We can see significantly strong patches of the BR asymmetry in the spatial BR asymmetry map on the top right, at $x$ = 27.54 Mm. The ${(\frac{B-R}{I})_m}$ value at the location of the black dot is -0.295, i.e., a redward spectral line asymmetry, and the spectrum on the bottom right panel also shows the same.
In Figure \ref{fig:brt46}, we have analyzed the cross-section of the cube at the same $x$ and $y$, but for $t$ = 920 s, we could see a blueward asymmetry with ${(\frac{B-R}{I})_m}$ value of 0.611. The secondary peak on the blue side of the spectra can be seen in the bottom right panel. Hence, an alternating trend of blueward and redward asymmetries can be seen with time. This is also depicted by the $x-t$ analysis in  Figure \ref{fig:brxt}.

\begin{figure*}[ht!]
    \centering

    \begin{minipage}{0.48\textwidth}
        \centering
        \begin{tikzpicture}
            \node[inner sep=0pt] (img1) {\includegraphics[width=\linewidth]{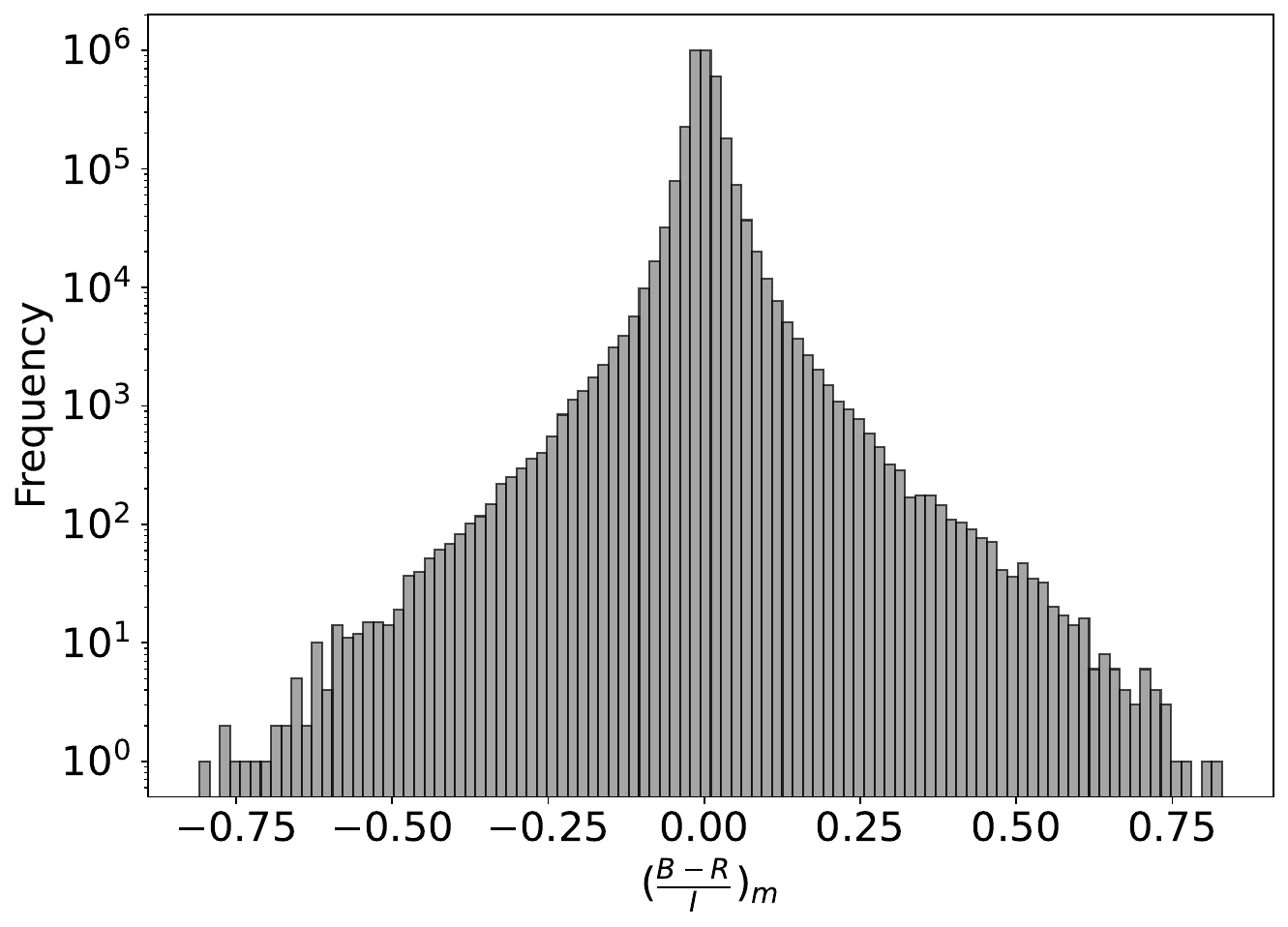}};
            \node[anchor=south west, xshift=-3pt, yshift=10pt] at (img1.north west) {\textbf{(a)}};
        \end{tikzpicture}
    \end{minipage}%
    \hfill
    \begin{minipage}{0.48\textwidth}
        \centering
        \begin{tikzpicture}
            \node[inner sep=0pt] (img2) {\includegraphics[width=\linewidth]{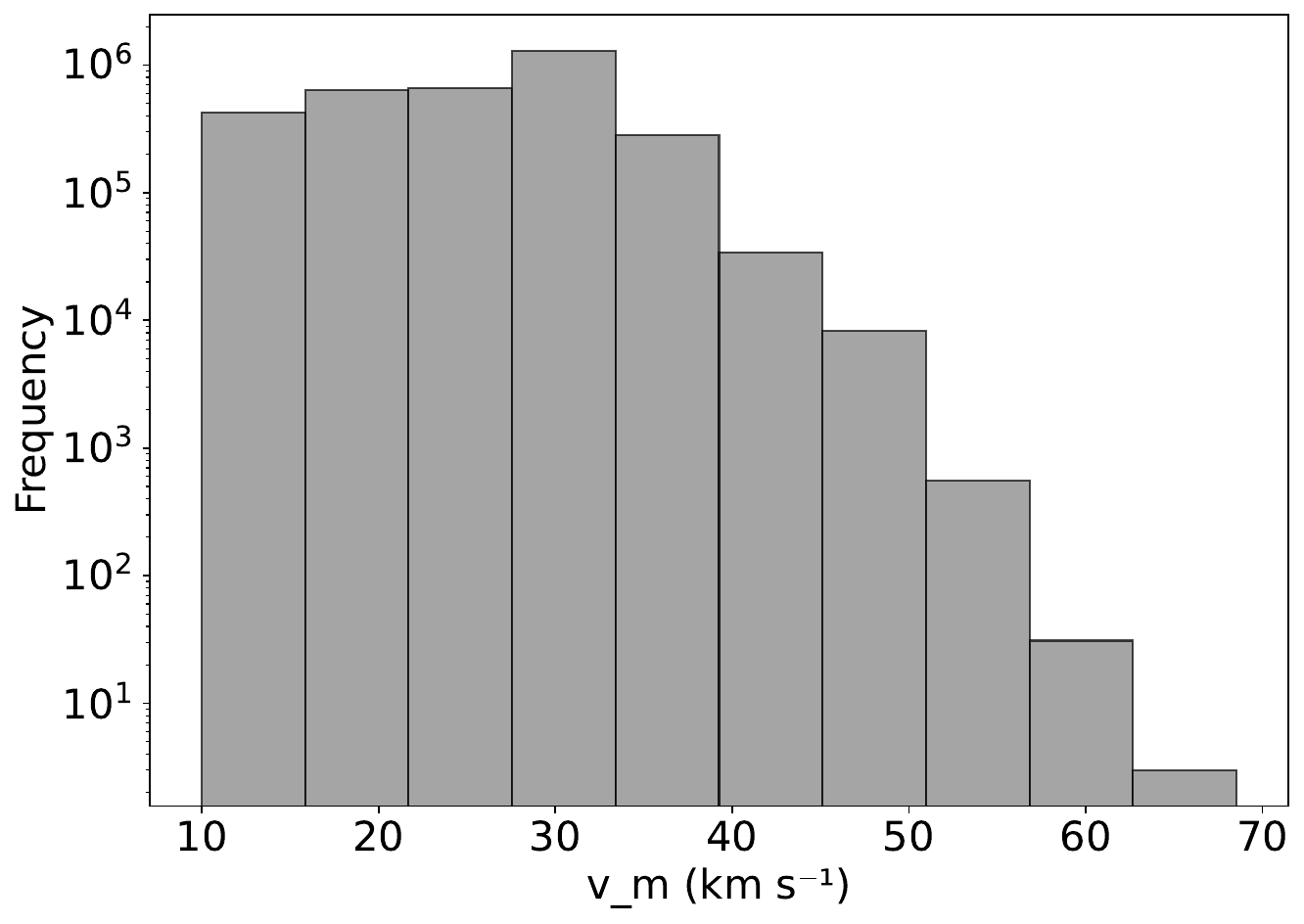}};
            \node[anchor=south west, xshift=-3pt, yshift=10pt] at (img2.north west) {\textbf{(b)}};
        \end{tikzpicture}
    \end{minipage}

    \vspace{0.6cm} 

    \begin{minipage}{0.48\textwidth}
        \centering
        \begin{tikzpicture}
            \node[inner sep=0pt] (img3) {\includegraphics[width=\linewidth]{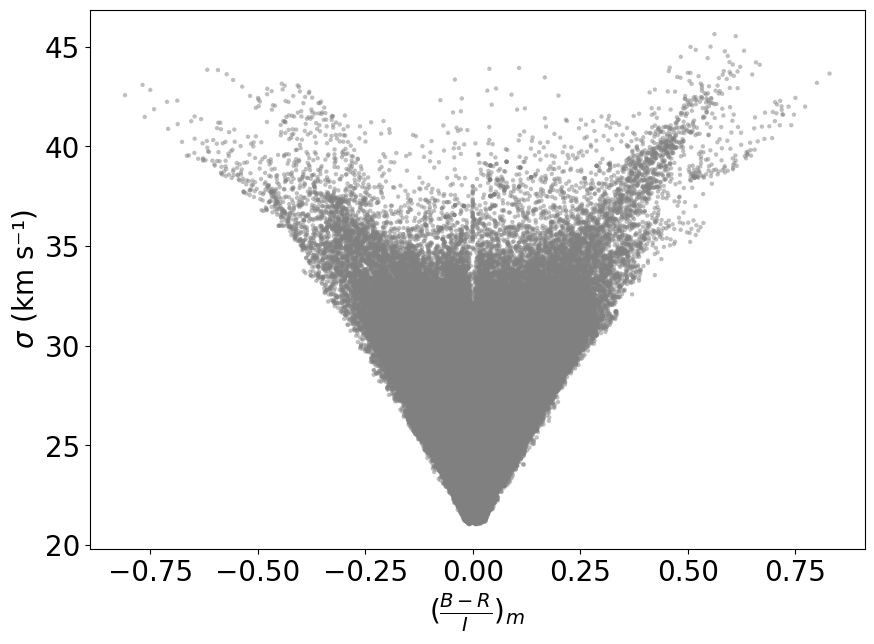}};
            \node[anchor=south west, xshift=-3pt, yshift=10pt] at (img3.north west) {\textbf{(c)}};
        \end{tikzpicture}
    \end{minipage}%
    \hfill
    \begin{minipage}{0.48\textwidth}
        \centering
        \begin{tikzpicture}
            \node[inner sep=0pt] (img4) {\includegraphics[width=\linewidth]{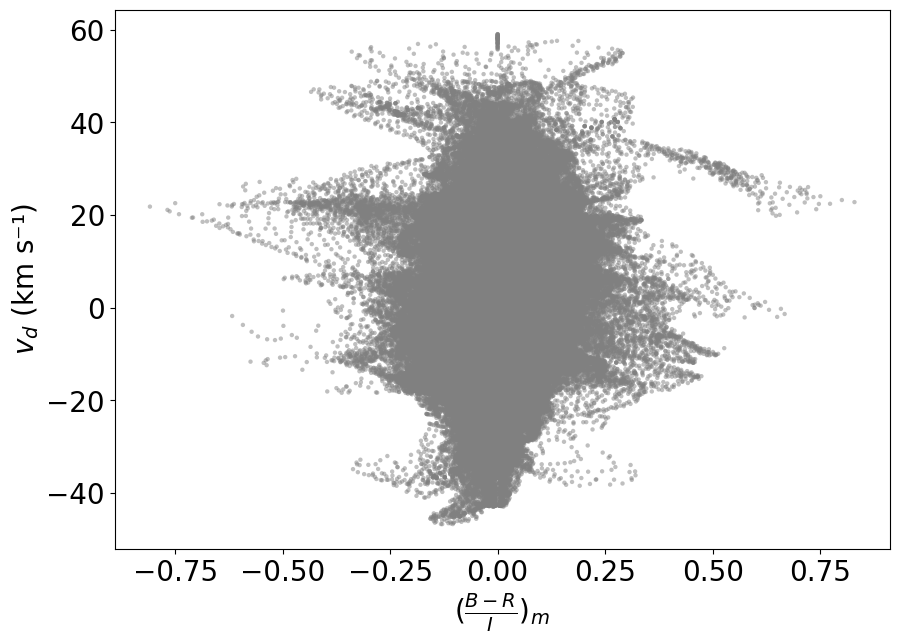}};
            \node[anchor=south west, xshift=-3pt, yshift=10pt] at (img4.north west) {\textbf{(d)}};
        \end{tikzpicture}
    \end{minipage}

    \vspace{0.6cm} 

    \caption{(a) Distribution of ${(\frac{B-R}{I})_m}$ ($y$-axis in log scale), 
    (b) distribution of $v_m$, the velocity corresponding to maximum asymmetry (secondary component), 
    (c) correlation between line width, $\sigma$, and normalized BR asymmetry, 
    (d) correlation between Doppler shifts, $v_d$, and normalized BR asymmetry, 
    for LOS integrated emission along the $z$-axis and driver velocity $v_{rms}$ = 15 km s$^{-1}$. 
    Line widths and Doppler shifts are derived from single Gaussian fits, while${(\frac{B-R}{I})_m}$ and $v_m$ are obtained from BR asymmetry analysis of the Fe XIII line at 10749 \AA.}
    \label{fig:br_st_l}
\end{figure*}

We have also plotted the distribution of ${(\frac{B-R}{I})_m}$, over the whole spatial domain and time of the simulation and found that this distribution is bimodal and nearly symmetric, which indicates the presence of both blue and red asymmetries, which can be noted in Figure \ref{fig:br_st_l}, panel (a). We found that the ${(\frac{B-R}{I})_m}$, values may reach up to around 75\% for the case when the rms value of the driver velocity is 15 km s$^{-1}$, which can be seen in panel (b). From the histogram of the velocities corresponding to ${(\frac{B-R}{I})_m}$, we found that they are distributed between 10-70 km s$^{-1}$ and peak at around 20-40 km s$^{-1}$. We also found that the line widths (FWHM), denoted by $\sigma$ in Figure \ref{fig:br_st_l}, panel (c), which were derived from the single Gaussian fit, increase with the increase in blueward and redward asymmetries and are in the range of 20-40 km s$^{-1}$. This correlation is because both spectral line asymmetries and increased line widths are due to the unresolved motions along the LOS. 
The Doppler shift, $v_d$, (derived from the single Gaussian fit ) vs BR asymmetry plot, is shown in panel (d) of Figure \ref{fig:br_st_l}, the Doppler shifts extend up to $\pm$50 km s$^{-1}$.

In our analysis, we find that for a blueshift (or redshift), both blueward and redward asymmetries are present, a behavior not reported in studies focusing on upflows \citep{2011tian}. Upflows and downflows are typically observed relative to a quasi-static background plasma, producing blue- and redshifts, respectively, together with asymmetries of the same sign. In contrast, for propagating transverse waves, the emission arises from multiple plasma elements of uniturbulent plasma with different densities, temperatures, and velocities, including both redshifted and blueshifted components along the LOS. At any given moment, some of these structures may move toward the observer while others move away. The emission from these inhomogeneous plasma elements, due to their distinct physical properties, contributes differently to the composite spectral profile. Consequently, even when the bulk plasma motion is blueward, a strong redshifted component can generate a redward asymmetry, and conversely, a strong blueshifted component within a predominantly redshifted plasma can produce a blueward asymmetry.
We have also carried out a similar BR asymmetry analysis for a higher rms driver velocity, i.e., 26 km s$^{-1}$, and found almost similar distributions; however, these findings are not included in the paper for the sake of simplicity.

For $v_{rms}$ = 26 km s$^{-1}$, we found that the ${(\frac{B-R}{I})_m}$ distribution is again bimodal but extends out to larger values because the driver's higher velocity leads to more turbulence and, hence, more asymmetry in the spectra. The distribution of velocities corresponding to ${(\frac{B-R}{I})_m}$ also peaks in the range of 30-40 km s$^{-1}$, but, in this case, the velocities at the higher end were more frequent compared to those of the case of driver velocity of 15 km s$^{-1}$.
The line widths were found to be higher than the previous case, in the 20-55 km s$^{-1}$ range. The Doppler shift was in the $\pm 60$ km s$^{-1}$ range, which is higher than the $v_{rms}$ = 15 km s$^{-1}$ case, which all accounts for the fact that the driver velocity is higher.

\subsection{Spectral line asymmetry along the random line of sights}
To mimic the real scenario of observations of the optically thin solar corona, we stack different realizations of the simulation to replicate the emission from multiple solar structures oscillating with different velocities and phases along the observer's LOS. Since, the structures in real observations are not always observed from the observer's LOS that is strictly perpendicular to or aligned with the loop axis, we randomly chose 100 segments corresponding to 21 different LOS across ($0\degree$, $15\degree$, $30\degree$, $45\degree$, $60\degree$, $75\degree$, $90\degree$, $105\degree$, $120\degree$, $135\degree$, $150\degree$, $165\degree$) and along ($45\degree$, $50\degree$, $55\degree$, $60\degree$, $65\degree$, $70\degree$, $75\degree$, $80\degree$, $85\degree$) the magnetic field direction and different time instants of the simulation. The emissions from these segments were integrated and averaged out. This case was simulated for a duration of 400 s, with a cadence of 20 s. 

\begin{figure*}[ht!]
    \centering

    \begin{minipage}{0.48\textwidth}
        \centering
        \begin{tikzpicture}
            \node[inner sep=0pt] (img1) {\includegraphics[width=\linewidth]{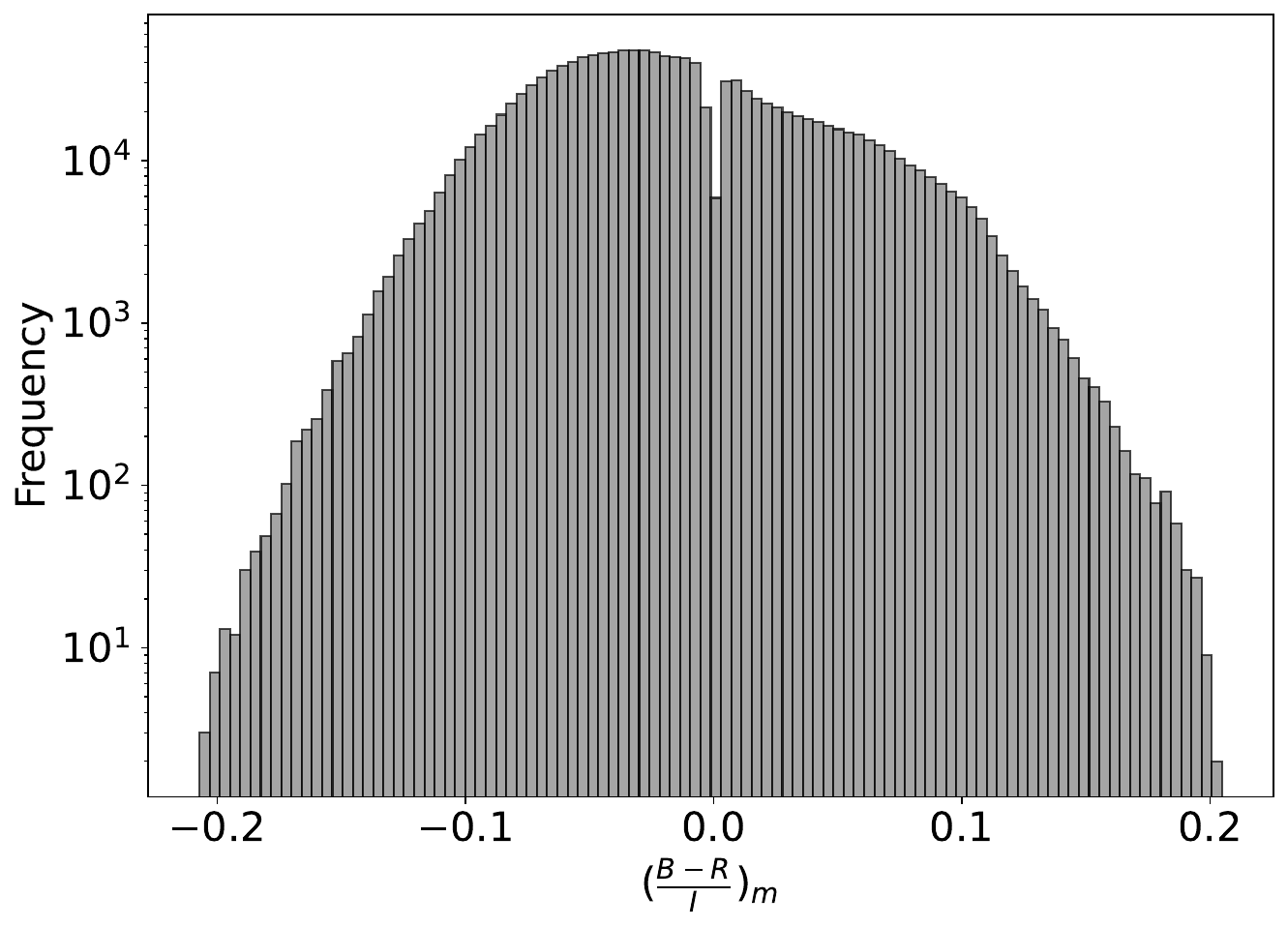}};
            \node[anchor=south west, xshift=-3pt, yshift=10pt] at (img1.north west) {\textbf{(a)}};
        \end{tikzpicture}
    \end{minipage}%
    \hfill
    \begin{minipage}{0.48\textwidth}
        \centering
        \begin{tikzpicture}
            \node[inner sep=0pt] (img2) {\includegraphics[width=\linewidth]{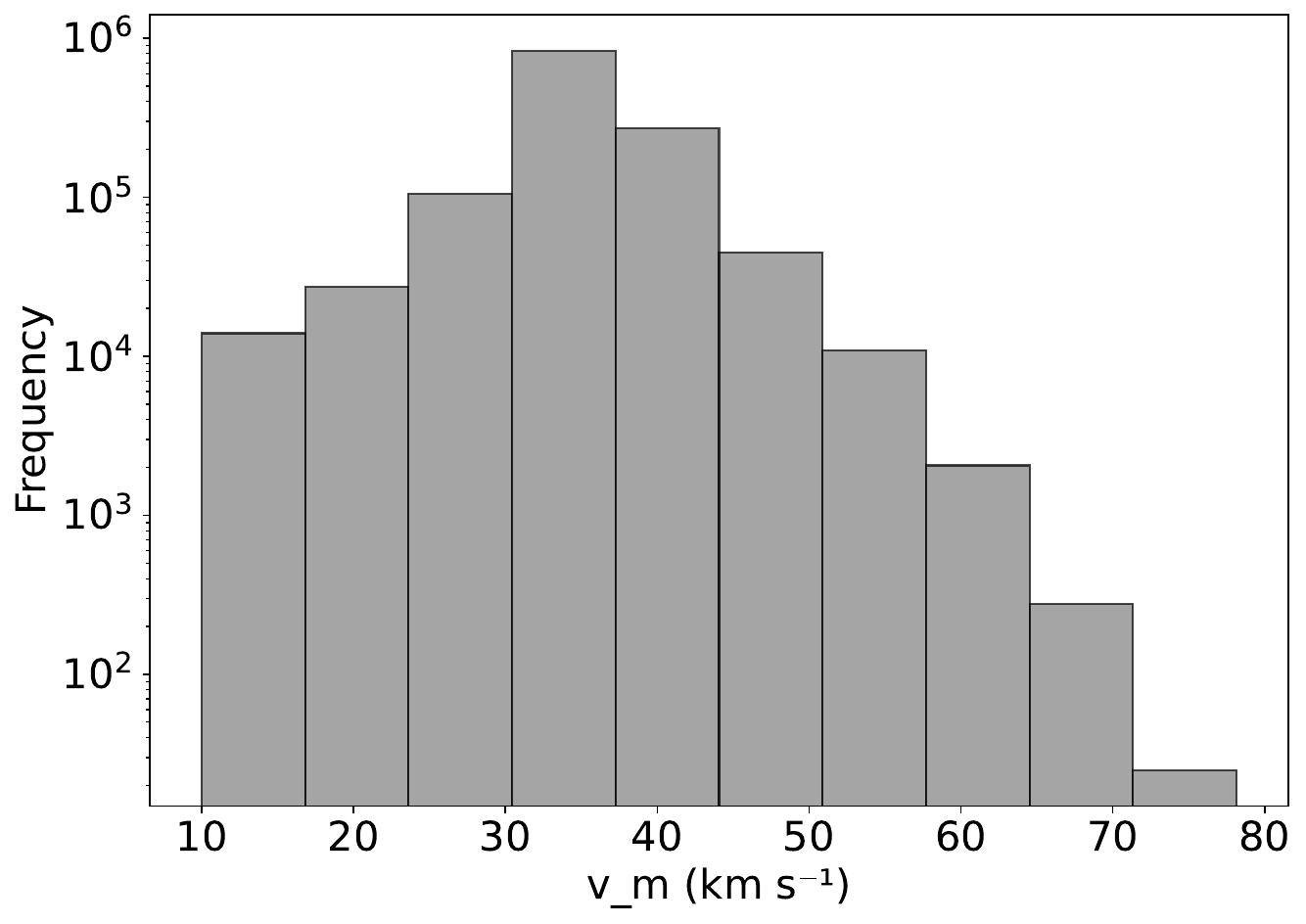}};
            \node[anchor=south west, xshift=-3pt, yshift=10pt] at (img2.north west) {\textbf{(b)}};
        \end{tikzpicture}
    \end{minipage}

    \vspace{0.6cm} 

    \begin{minipage}{0.48\textwidth}
        \centering
        \begin{tikzpicture}
            \node[inner sep=0pt] (img3) {\includegraphics[width=\linewidth]{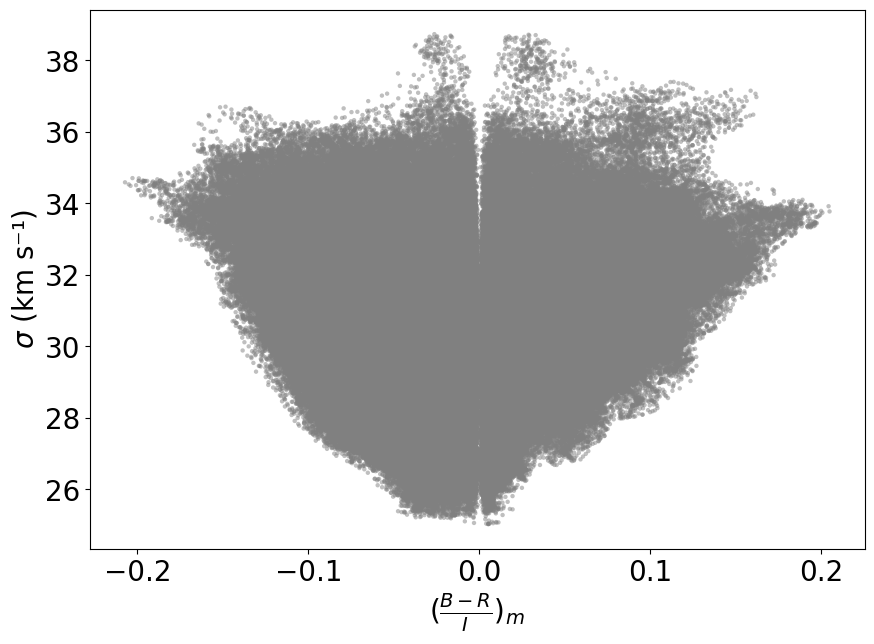}};
            \node[anchor=south west, xshift=-3pt, yshift=10pt] at (img3.north west) {\textbf{(c)}};
        \end{tikzpicture}
    \end{minipage}%
    \hfill
    \begin{minipage}{0.48\textwidth}
        \centering
        \begin{tikzpicture}
            \node[inner sep=0pt] (img4) {\includegraphics[width=\linewidth]{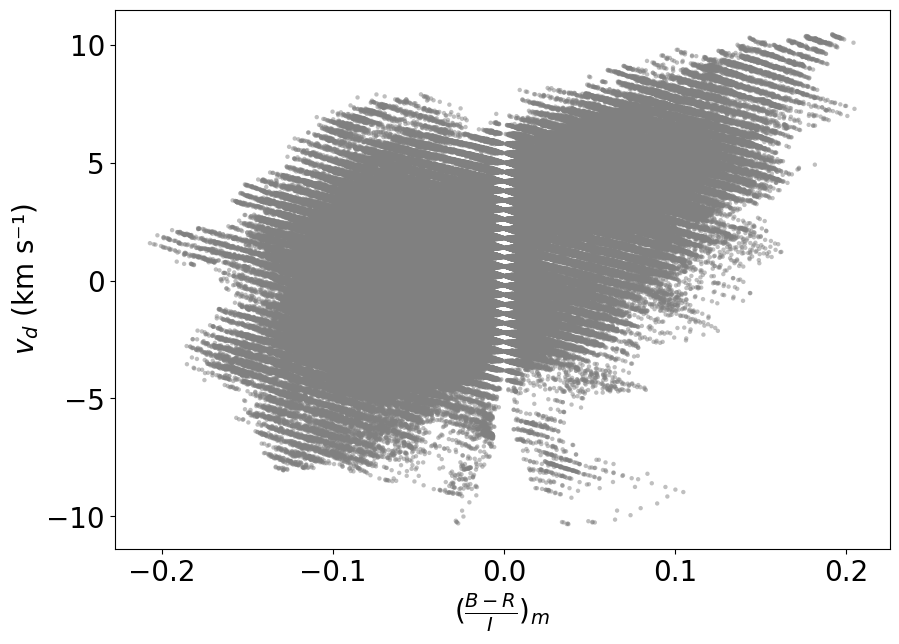}};
            \node[anchor=south west, xshift=-3pt, yshift=10pt] at (img4.north west) {\textbf{(d)}};
        \end{tikzpicture}
    \end{minipage}

    \vspace{0.6cm} 

    \caption{Same as Figure \ref{fig:br_st_l} but for emission integrated and combined for different random LOS along and across the guiding magnetic field direction, having the same driver velocity, $v_{rms}$ = 15 km s$^{-1}$, }
    \label{fig:br_st_3}
\end{figure*}

An analysis of this data for $v_{rms}$ = 15 km s$^{-1}$ shows that the bimodal distribution of ${(\frac{B-R}{I})_m}$ shrunk to lower values. From Figure \ref{fig:br_st_3}, panel(a), it could be seen that the ${(\frac{B-R}{I})_m}$ value is now reduced to 20 \%, in contrast with 75 \% as computed for a single LOS of the forward modelled cube. This could be attributed to the summation of various asymmetric spectra (obtained from various single LOS integrations of the forward modelled cube), leading to a more symmetric spectrum. Although the bimodal nature of BR asymmetries is preserved in this case, a slight variation in the symmetry of this distribution can be seen due to the randomness of polarization of oscillating features in the integrated emission. 
The velocity of the secondary component peaks at 30-40 km s$^{-1}$. The line width shows a similar expected increasing trend with the BR asymmetry, and the line widths fall in the range of 25-38 km s$^{-1}$, and the Doppler shift values extend to $\pm 10$ km s$^{-1}$.In panel (d) of Figure \ref{fig:br_st_3}, blueward asymmetry seems to correlate with redshifts, while redward asymmetry seems to correlate with blueshifts. However, we would like to point out that this is not a universal behavior for cases involving multiple structures along the observer's LOS; rather, it reflects one particular realization where the random selection of oscillation phases and polarizations produced such a correlation. For other random configurations, we found that the correlation differs for different cases; however, we found that the distribution extends for all four quadrants for all cases.  

\begin{figure}[ht!]
    \centering

    \begin{minipage}[t]{0.48\textwidth}
        \centering
        \begin{tikzpicture}
            \node[inner sep=0pt] (img1) {\includegraphics[width=9cm,height=10cm]{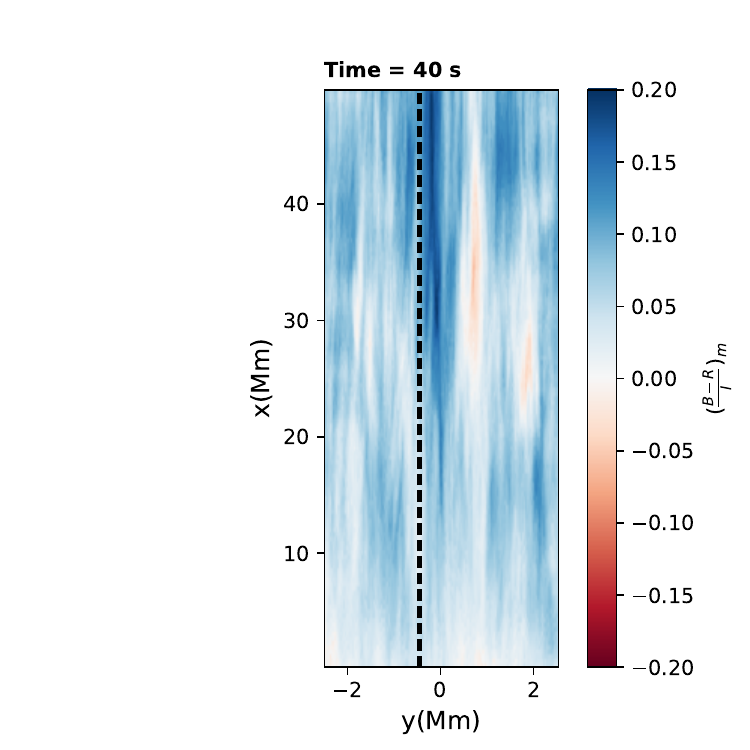}};
            \node[anchor=south west, xshift=-3pt, yshift=10pt] at (img1.north west) {\textbf{(a)}};
        \end{tikzpicture}
    \end{minipage}%
    \hfill
    \begin{minipage}[t]{0.45\textwidth}
        \centering
        \begin{tikzpicture}
            \node[inner sep=0pt] (img2) {\includegraphics[width=\linewidth]{figure10b.pdf}};
            \node[anchor=south west, xshift=-3pt, yshift=10pt] at (img2.north west) {\textbf{(b)}};
        \end{tikzpicture}
    \end{minipage}

    \vspace{0.1cm} 

    \begin{minipage}[t]{0.8\textwidth}
        \centering
        \begin{tikzpicture}
            \node[inner sep=0pt] (img3) {\includegraphics[width=\linewidth,height=10cm]{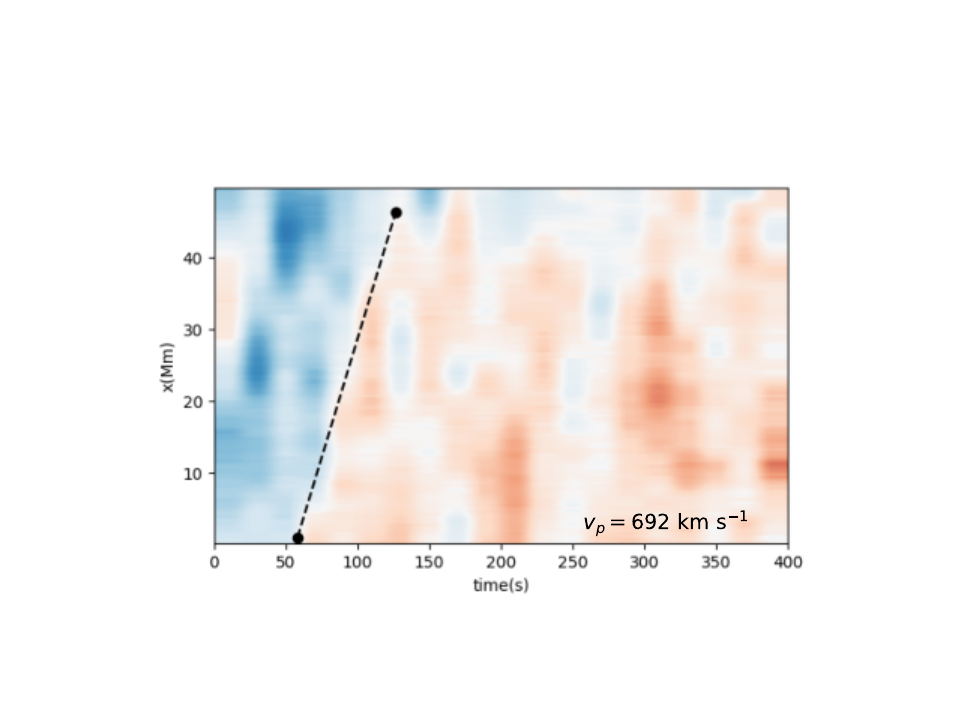}};
            \node[anchor=south west, xshift=-3pt, yshift=10pt] at (img3.north west) {\textbf{(c)}};
        \end{tikzpicture}
    \end{minipage}

    \caption{Distance-time maps for the BR asymmetry features for the case shown in Figure \ref{fig:br_st_l}, where $v_{rms}$ = 15 km s$^{-1}$ and emission is integrated and combined for different random LOS along and across the guiding magnetic field direction:  
    (a) slit position is depicted by the dashed line on the BR asymmetry map at $t$= 40 s;  
    (b) distribution of speeds of propagation of the features in the BR asymmetry maps, represented by $v_p$;  
    (c) distance-time map of the features along the slit length in (a), where black dots mark the edges of BR asymmetry patches, and the dashed line joining them represents the slope used to calculate the propagation speeds.}
    \label{fig:xt1}
\end{figure}

In Figure \ref{fig:xt1}, panel (c), we have plotted the $x-t$ map for a spatial BR asymmetry map for the case of randomly polarized emission with driver velocity of $v_{rms}$ = 15 km s$^{-1}$.
A 5-pixel wide artificial slit was kept at $y=$ -0.48 Mm, 
and the $x-t$ map for the region covered by the slit is plotted; here, the dashed line represents the location of the artificial slit. To estimate the values of the propagation speeds of features present in this $x-t$ map, we chose two endpoints along the length of these features and calculated the slope (depicted by the dashed line in the $x-t$ map). We made several $x-t$ maps by adjusting the slit positions along the $y-$axis and estimated the speeds for different features present in these $x-t$ maps. The histogram for the distribution of these speeds is shown in panel (b) of Figure \ref{fig:xt1}. The $x-t$ map in panel (c) displays the reversal of BR asymmetry between blue and red over time. The histogram shows prominent peaks at values of $v_p$ around 700--800~km~s$^{-1}$. Notably, this range matches the phase velocity of the driven waves ($\approx 800$~km~s$^{-1}$), which we previously estimated from Doppler velocity maps but have omitted here for simplicity.
 The observed spectral line asymmetries are the manifestation of the turbulence developed by these waves in the inhomogeneous solar plasma and are also due to the observer's LOS superposition of emissions from multiple structures oscillating with different phases and polarizations.

\subsection{Observational aspects of the transverse wave-induced spectral line asymmetries: DKIST}
\begin{figure*}[ht!]
    \centering
    \begin{interactive}{animation}{figure11.mp4}
    \includegraphics[width=\textwidth, keepaspectratio]{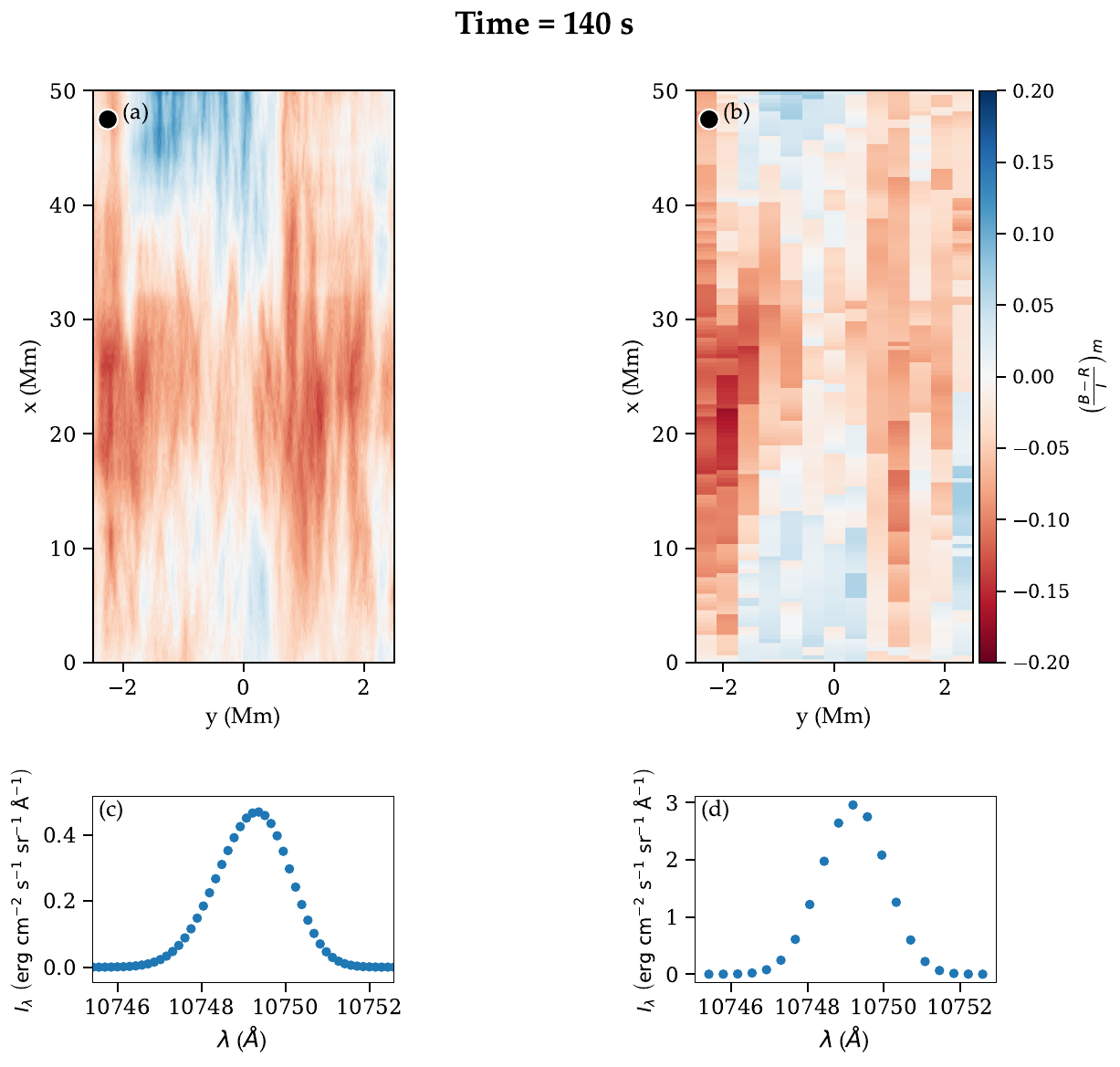} 
        
    \end{interactive}
    
    \caption{BR asymmetry maps for emission integrated along different random LOS across and along the magnetic field direction, $v_{rms}$ = 15 km s$^{-1}$, forward modeled for the DKIST at 10749 \AA:  
    (a) Spatial BR asymmetry map at the resolution of the simulation at $t$= 140 s;  
    (b) Same map at the spatial and spectral resolution of the Cryo-NIRSP/DKIST;  
    (c) Spectrum at the location of the black dot in (a) showing $\sim$20\% BR asymmetry towards the blue wing;  
    (d) Spectrum at a similar location at the Cryo-NIRSP/DKIST resolution, marked by the black dot in (b). A full animation of the evolution of BR asymmetry maps for the random LOS case, along with corresponding DKIST-resolution maps, is provided in the supplementary materials.}
    \label{fig:brmaps_dkist}
\end{figure*}

\begin{figure*}[ht!]
    \centering

    \begin{minipage}{0.48\textwidth}
        \centering
        \begin{tikzpicture}
            \node[inner sep=0pt] (img1) {\includegraphics[width=\linewidth]{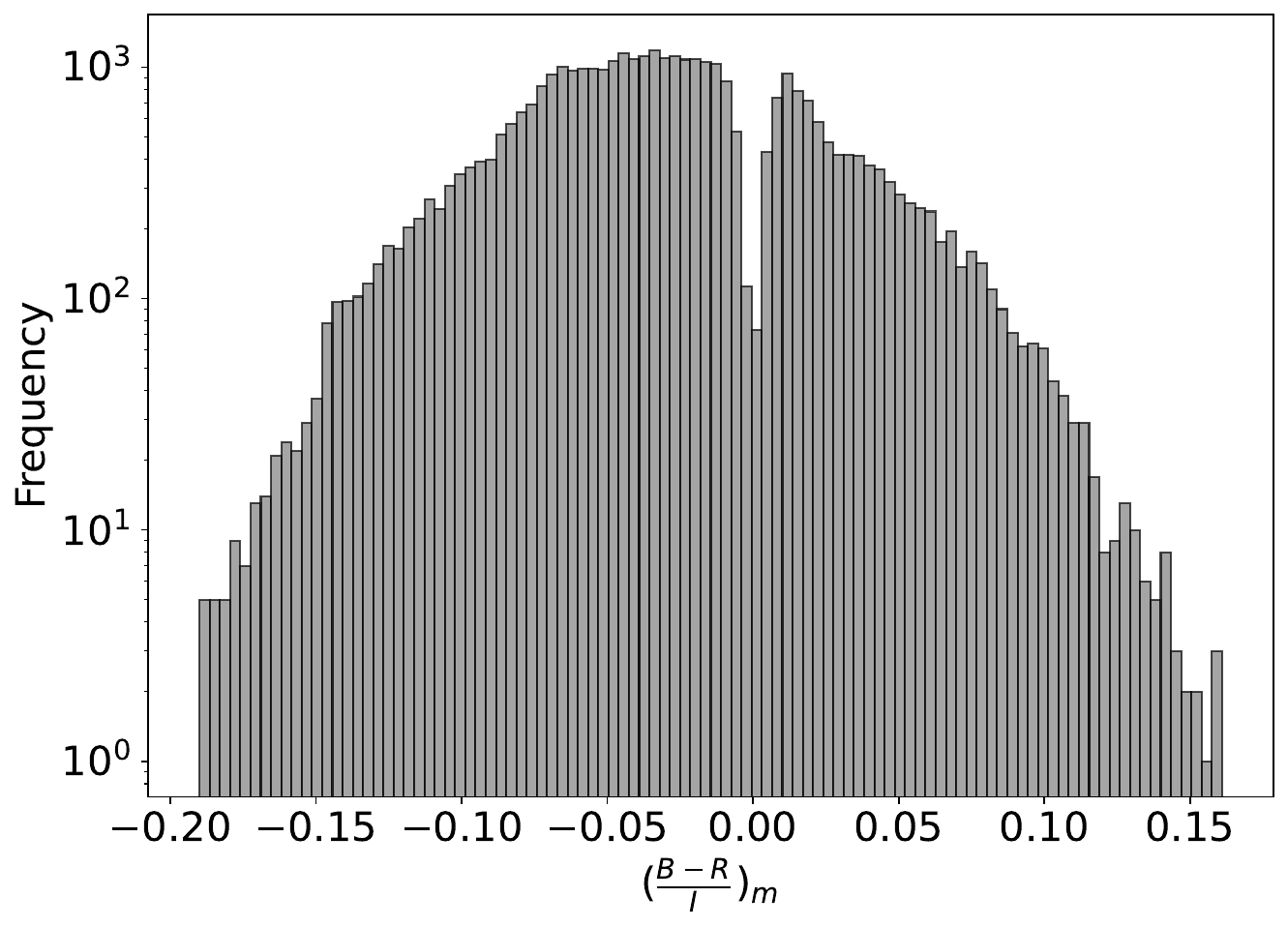}};
            \node[anchor=south west, xshift=-3pt, yshift=10pt] at (img1.north west) {\textbf{(a)}};
        \end{tikzpicture}
    \end{minipage}%
    \hfill
    \begin{minipage}{0.48\textwidth}
        \centering
        \begin{tikzpicture}
            \node[inner sep=0pt] (img2) {\includegraphics[width=\linewidth]{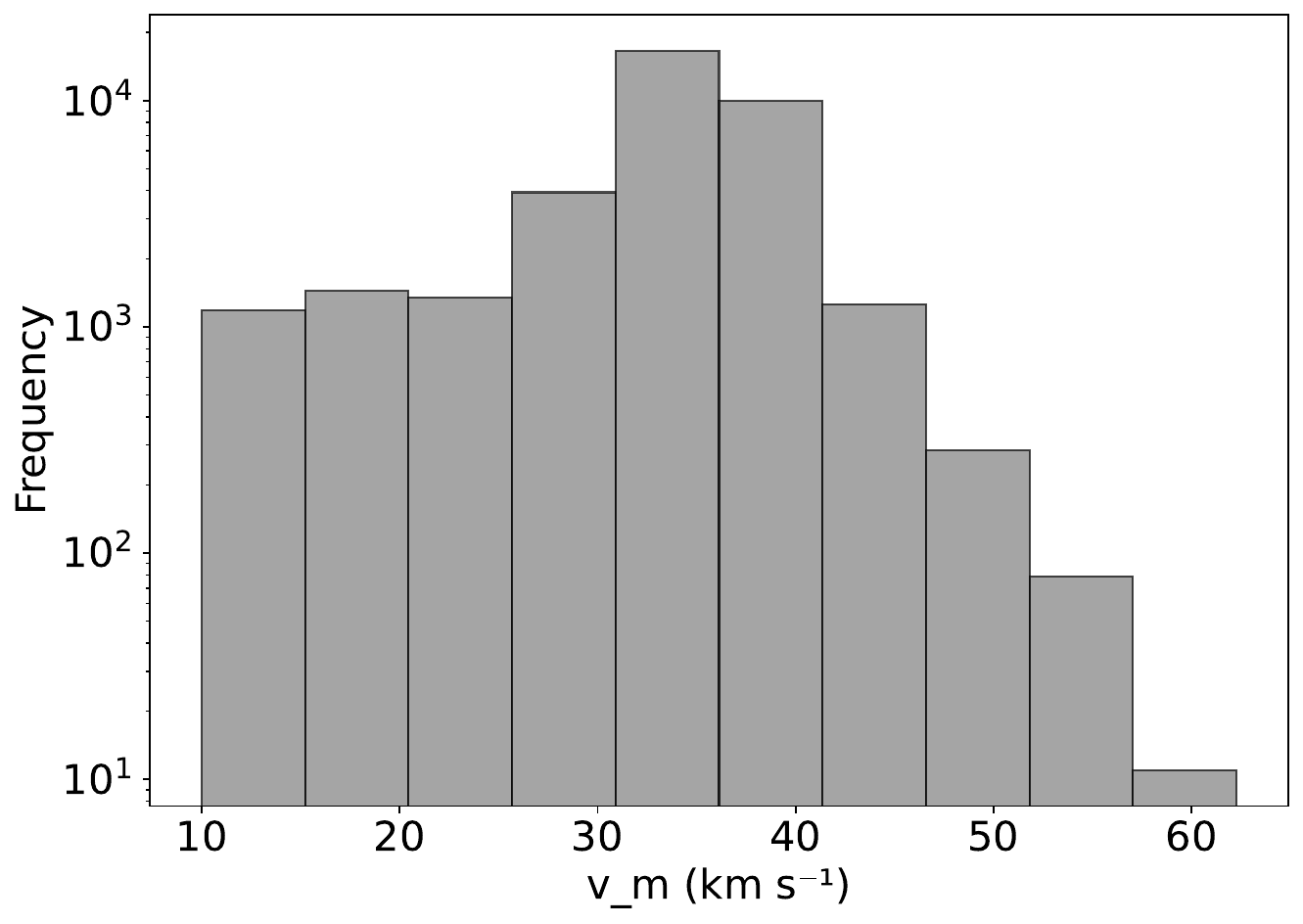}};
            \node[anchor=south west, xshift=-3pt, yshift=10pt] at (img2.north west) {\textbf{(b)}};
        \end{tikzpicture}
    \end{minipage}

    \vspace{0.6cm} 

    \begin{minipage}{0.48\textwidth}
        \centering
        \begin{tikzpicture}
            \node[inner sep=0pt] (img3) {\includegraphics[width=\linewidth]{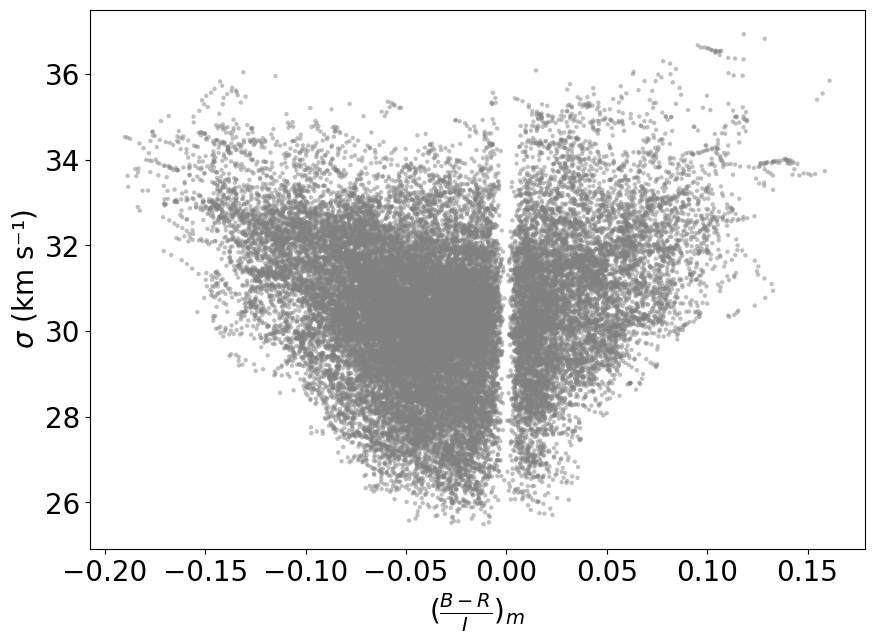}};
            \node[anchor=south west, xshift=-3pt, yshift=10pt] at (img3.north west) {\textbf{(c)}};
        \end{tikzpicture}
    \end{minipage}%
    \hfill
    \begin{minipage}{0.48\textwidth}
        \centering
        \begin{tikzpicture}
            \node[inner sep=0pt] (img4) {\includegraphics[width=\linewidth]{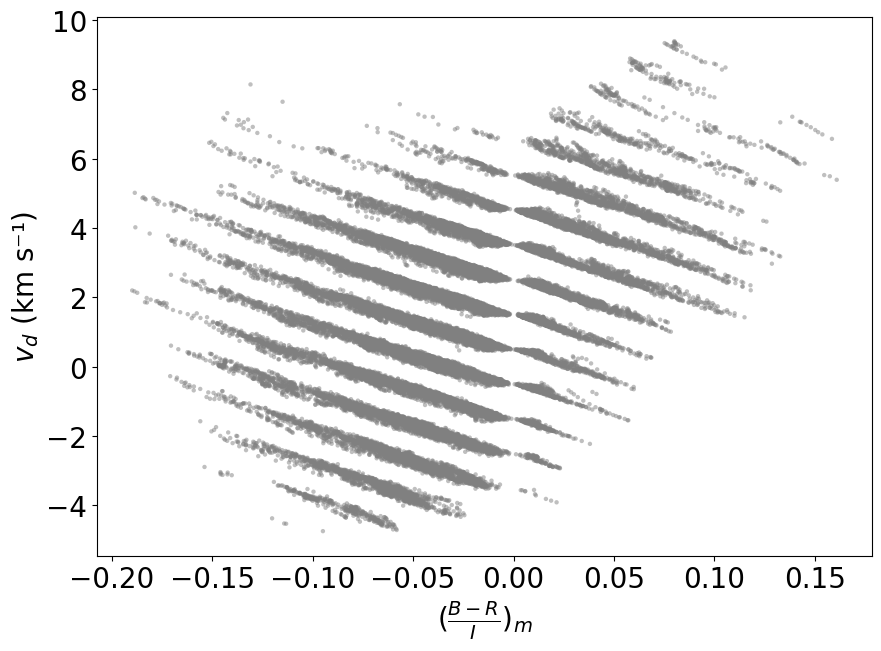}};
            \node[anchor=south west, xshift=-3pt, yshift=10pt] at (img4.north west) {\textbf{(d)}};
        \end{tikzpicture}
    \end{minipage}

    \vspace{0.6cm} 

    \caption{Same as Figure \ref{fig:br_st_l} but for the driver velocity $v_{rms}$ = 15 km s$^{-1}$, for emission integrated along different random LOS across and along the magnetic field direction, forward modeled for the Cryo-NIRSP/DKIST spectral and spatial resolution at 10749 \AA.}
    \label{fig:br_st_4}
\end{figure*}
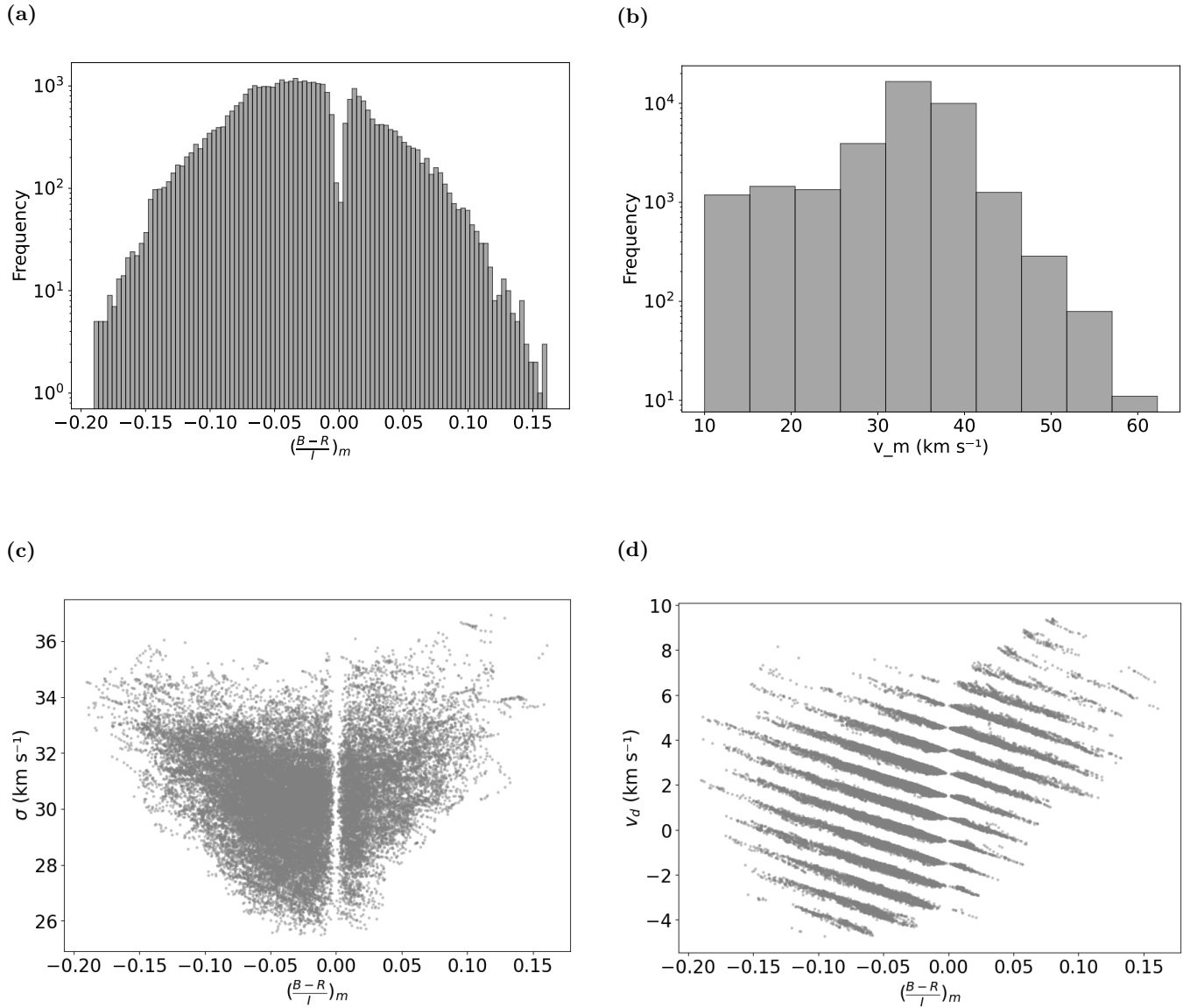

\begin{figure}[ht!]
    \centering

    \begin{minipage}[t]{0.48\textwidth}
        \centering
        \begin{tikzpicture}
            \node[inner sep=0pt] (img1) {\includegraphics[width=9cm,height=10cm]{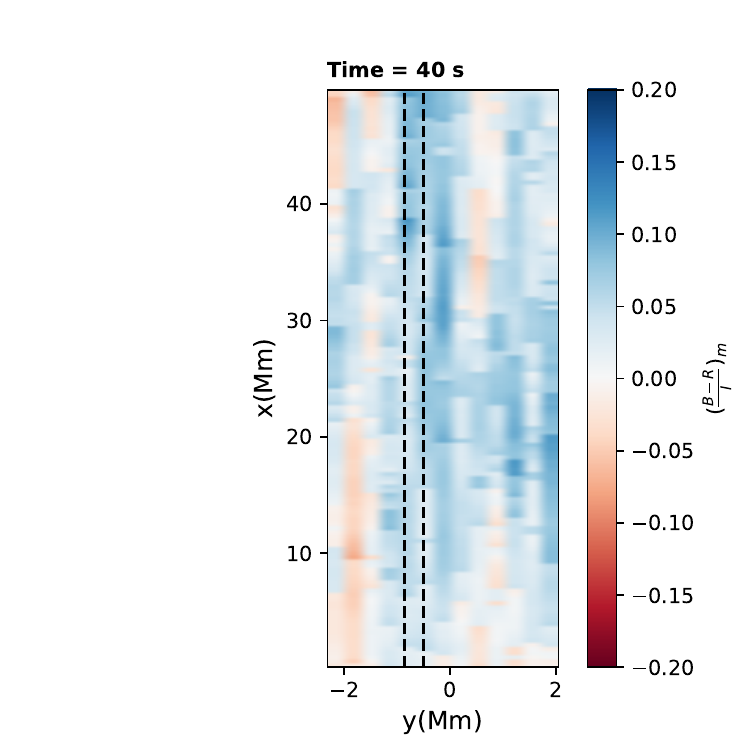}};
            \node[anchor=south west, xshift=-3pt, yshift=10pt] at (img1.north west) {\textbf{(a)}};
        \end{tikzpicture}
    \end{minipage}%
    \hfill
    \begin{minipage}[t]{0.45\textwidth}
        \centering
        \begin{tikzpicture}
            \node[inner sep=0pt] (img2) {\includegraphics[width=\linewidth]{figure13b.pdf}};
            \node[anchor=south west, xshift=-3pt, yshift=10pt] at (img2.north west) {\textbf{(b)}};
        \end{tikzpicture}
    \end{minipage}

    \vspace{1cm} 

    \begin{minipage}[t]{0.8\textwidth}
        \centering
        \begin{tikzpicture}
            \node[inner sep=0pt] (img3) {\includegraphics[width=\linewidth,height=10cm]{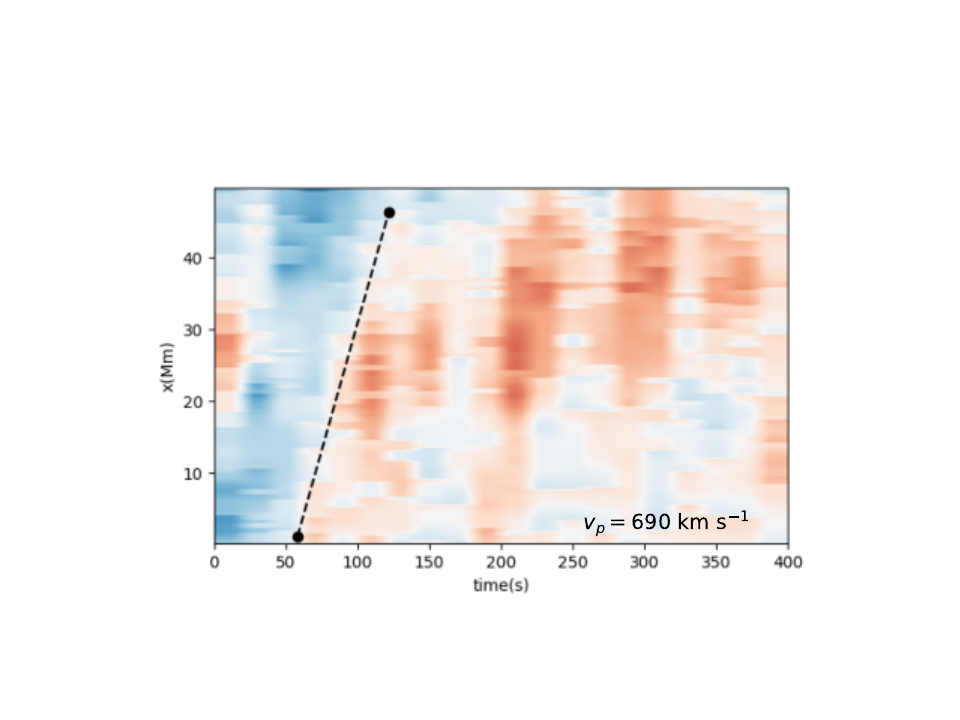}};
            \node[anchor=south west, xshift=-3pt, yshift=10pt] at (img3.north west) {\textbf{(c)}};
        \end{tikzpicture}
    \end{minipage}

    \vspace{0.6cm} 

    \caption{Same as Figure \ref{fig:xt1} but for distance-time maps as observed by the Cryo-NIRSP/DKIST, here $v_{rms}$ = 15 km s$^{-1}$.}
    \label{fig:xt2}
\end{figure}

In our study, we could observe the asymmetries in spectral lines due to our simulation's high spatial and spectral resolution, which was the limitation of previous solar spectrographs. However, the advent of recent and upcoming instruments presents the opportunity to observe and study the spectral line asymmetries caused by these propagating transverse waves. Here, we have forward-modelled our data to study what BR asymmetry signatures will be observed if we consider the capabilities of the Cryogenic Near Infrared Spectropolarimeter (Cryo-NIRSP) of DKIST \citep{2020SoPh..295..172R}. Cryo-NIRSP is a slit-based spectrograph and context imaging instrument, the spatial sampling for Cryo-NIRSP is 0.5'' of low-resolution slit, and spectral resolution is around 30,000 for coronal observations \citep{2023SoPh..298....5F}.
We have coarsened our simulation's spatial and spectral resolution with the DKIST's resolution. We have considered the same dataset as Figure \ref{fig:br_st_3}, which is a more realistic case.  
We have used the DataArray.coarsen() function of the Xarray \footnote{\url{https://xarray.dev}} library of Python \citep{hoyer2017xarray}, to downsample the data along the $y-$axis. Neighboring data points were grouped in the $y$ direction and summed over the window to get the spatial resolution of Cryo-NIRSP, while the spectral resolution was reduced by forward modeling using the FoMo tool. In Figure \ref{fig:brmaps_dkist}, on the left panel, we have shown a BR asymmetry map for the case of random integration of different LOS with v$_{rms}$ =15 km s$^{-1}$. On the right panel is the same map degraded with the DKIST resolution. We can see that the important BR asymmetry features are preserved even though very fine details may get lost in the case of the DKIST. Hence, we may be able to observe spectral line asymmetries caused by the propagating transverse waves from the DKIST. We have also performed a statistical analysis similar to the previous cases. In Figure \ref{fig:br_st_4}, we can see that BR asymmetry and velocity distributions are very similar to Figure \ref{fig:br_st_3}, with velocity of asymmetric component peaking at 30-40 km s$^{-1}$. The line width is 25-35 km s$^{-1}$.
The Doppler shift velocities are between +4 to -10 km s$^{-1}$. The gaps(pixelation) in this scatter plot are due to the data coarsening because of the averaging of Doppler shifts over many pixels, which give rise to quantization effects limiting the values of Doppler shifts and BR asymmetry to correlate only for specific ranges. Also, we can note an uneven distribution of BR asymmetries, with the excess presence of redward asymmetries compared to blueward asymmetries, which is only due to the random selection of various LOS, hence the opposite scenerio of excess presence of blueward spectral line asymmetry is also possible.
A similar result as in Figure \ref{fig:xt1} is obtained for the $x-t$ maps in Figure \ref{fig:xt2} for the emission degraded to the DKIST resolution. Here the slit width is 1 pixel wide, which is represented by two dashed lines in the top left panel of the Figure \ref{fig:xt2}. We can note the distribution of the speeds of features in the $x-t$ map are distributed around 600-900 km s$^{-1}$. We found that the forward-modeled results for the DKIST case agree with those of simulations. From this analysis we propose the possibility of observing these signatures of BR asymmetry due to the propagating transverse waves using Cryo-Nirsp.

\section{Discussion and Conclusion} \label{sec:cite}

Our analysis demonstrates that uniturbulence, previously shown to arise from unidirectionally propagating transverse MHD waves in inhomogeneous plasma, together with LOS superposition of oscillating structures, produces spectral line asymmetries through the unequal contributions of plasma elements with different densities, temperatures, and velocities.
From our study, a distinctive trait of the asymmetries induced by propagating transverse waves is found to be their nearly symmetric bimodal distribution, with both redward and blueward asymmetries present and alternating with time and height. Our statistical analysis further reveals that the magnitude of the BR asymmetry, $(\frac{B-R}{I})_m$, can reach values as high as 75\%. Additionally, the peak velocities of the secondary components associated with these asymmetries are typically below 100 km s$^{-1}$, in contrast to previous studies of plasma upflows, which reported values in the range of 50--150 km s$^{-1}$. However, in real observations, the optically thin nature of the solar corona results in the simultaneous detection of emissions from multiple structures oscillating with different phases and polarizations along the line of sight (LOS). This superposition introduces randomness, leading to an asymmetric bimodal distribution in which one type of asymmetry tends to dominate over the other. To investigate this effect, we analyzed synthetic spectra obtained by combining emissions from random LOS and different time instances, thereby mimicking solar structures oscillating with varying phases and velocities. The corresponding statistical analysis revealed that $(\frac{B-R}{I})_m$ in this scenario reaches only up to 20\%, a value comparable to that reported for plasma upflows, which typically ranges from a few percent to about 30\% of the primary emission \citep{2011tian, 2010A&A...521A..51P}. However, the velocities of the secondary components in our case still remain below 100 km s$^{-1}$, in contrast to the higher secondary component velocities typically associated with plasma upflows. For the random LOS integration, the Doppler shifts are on the order of 10 km s$^{-1}$, with Gaussian line widths of approximately 40 km s$^{-1}$. The correlation between Doppler shifts and BR asymmetries extends across all four quadrants, reflecting contributions from plasma elements moving in both directions along the LOS. This behavior contrasts with upflows, where only blueshifts correlate with blueward asymmetries.


An important distinction between these two types of asymmetries lies in their strong dependence on the LOS angle. In the case of upflow-related asymmetries, previous studies focused on on-disc coronal loops, where plasma flows are aligned with the loop axis \citep{2011tian,2009ApJ...701L...1D,2011ApJ...732...84M}, resulting in maximum asymmetries observed along the loops. In contrast, our study examines off-limb coronal plumes, where the asymmetries arise from turbulence generated by perpendicular inhomogeneities and propagating transverse waves. Consequently, the strongest asymmetries are observed across the loop, i.e., along LOS orientations perpendicular to the loop axis.  Since the observed asymmetries are driven by propagating transverse waves, they are expected to propagate along the height of the structure. By performing time-distance analysis of patch-like features in the BR asymmetry maps, we measured propagation velocities in the range of 600–1000 km s$^{-1}$, consistent with the phase velocity of $\sim$ 800 km s$^{-1}$ of the transverse waves. The limited spectral, spatial, and temporal resolution of earlier spectrographs likely concealed these signatures in past observations. However, modern and upcoming instruments such as DKIST and MUSE offer the capability to detect them. To test this, we degraded our synthetic data to the spectral and spatial resolution of Cryo-NIRSP/DKIST and found that the signatures of bimodal BR asymmetry remain detectable, with statistical properties consistent with those of the original numerical results. Our model does not include non-ideal effects such as wave damping, thermal conduction, and viscosity, which may influence the properties of the resulting asymmetries. In our model, the structure is considered only up to a height of $50\,\mathrm{Mm}$, where damping effects are not expected to play a significant role. At greater heights, however, such effects may become non-negligible and could influence the results concerning BR asymmetry. Incorporating these effects will be an important direction for future work.

\nolinenumbers 
\section{acknowledgments}
\nolinenumbers
We thank the anonymous referee for their constructive comments and insightful suggestions, which greatly improved the quality and clarity of this paper.
A.S. is supported by the funds from the
Department of Science and Technology (DST), Government
of India, through Aryabhatta Research Institute of
Observational Sciences (ARIES), India. V.P.
is supported by the SERB start-up research grant (File no. SRG/2022/001687).
TVD was supported by the C1 grant TRACEspace of Internal Funds KU Leuven and a Senior Research Project (G088021N) of the FWO Vlaanderen. Furthermore, TVD received financial support from the Flemish Government under the long-term structural Methusalem funding program, project SOUL:
Stellar evolution in full glory, grant METH/24/012 at KU Leuven. The research that led to these results was subsidised by the Belgian Federal Science Policy Office through the contract B2/223/P1/CLOSE-UP. It is also part of the DynaSun project and has thus received funding under the Horizon Europe programme of the European Union under grant agreement (no. 101131534). Views and opinions expressed are however those of the author(s) only and do not necessarily reflect those of the European Union and therefore the European Union cannot be held responsible for them. The authors acknowledge ChatGPT (OpenAI) and Grammarly for language refinement and paraphrasing support.

\appendix \label{appendix}

\section{Random lines of sight integration for $v_{rms}$ = 26 
km\,s$^{-1}$}

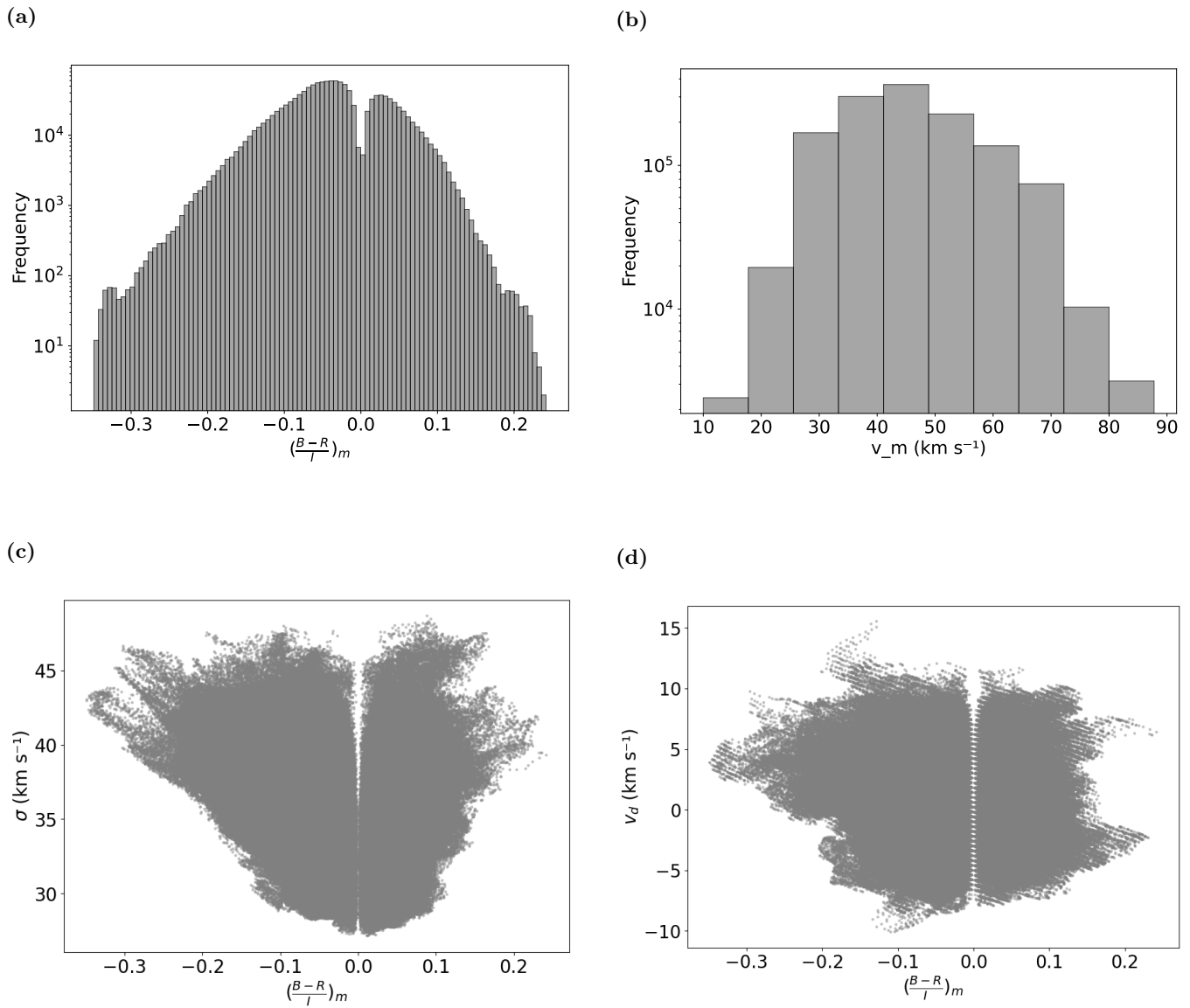
\begin{figure*}[ht!]
    \centering

    \begin{minipage}{0.48\textwidth}
        \centering
        \begin{tikzpicture}
            \node[inner sep=0pt] (img1) {\includegraphics[width=\linewidth]{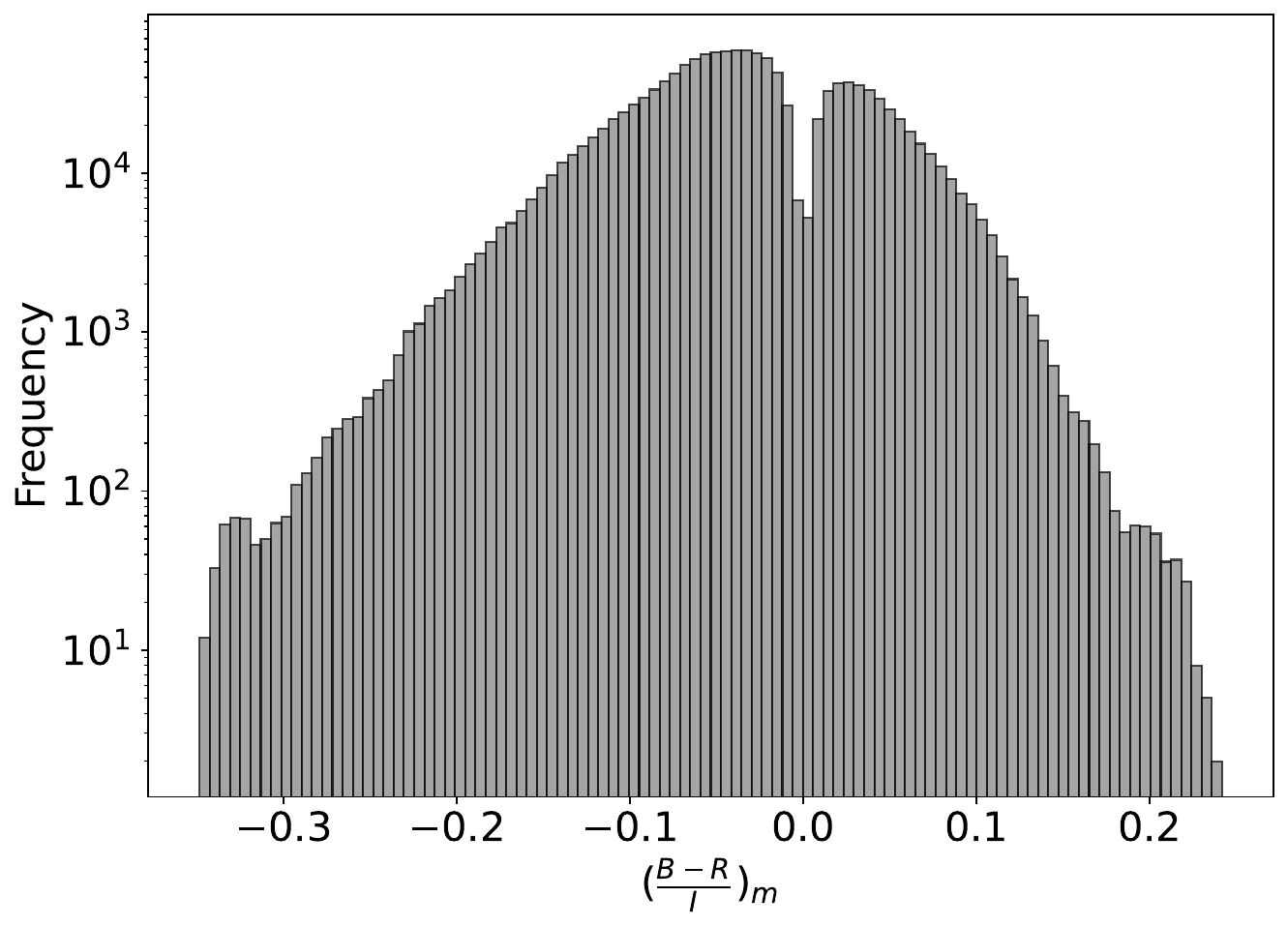}};
            \node[anchor=south west, xshift=-3pt, yshift=10pt] at (img1.north west) {\textbf{(a)}};
        \end{tikzpicture}
    \end{minipage}%
    \hfill
    \begin{minipage}{0.48\textwidth}
        \centering
        \begin{tikzpicture}
            \node[inner sep=0pt] (img2) {\includegraphics[width=\linewidth]{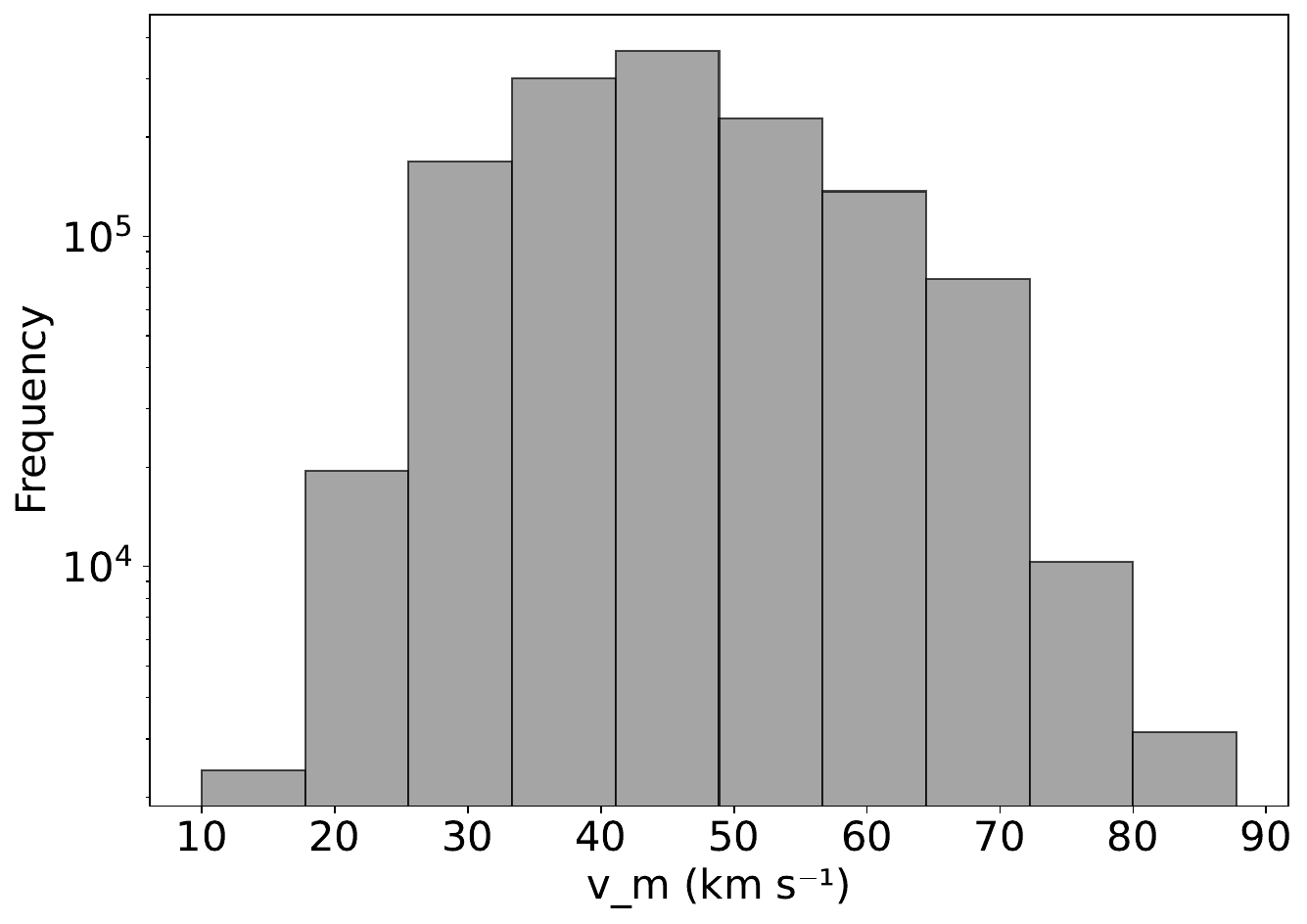}};
            \node[anchor=south west, xshift=-3pt, yshift=10pt] at (img2.north west) {\textbf{(b)}};
        \end{tikzpicture}
    \end{minipage}

    \vspace{0.6cm} 

    \begin{minipage}{0.48\textwidth}
        \centering
        \begin{tikzpicture}
            \node[inner sep=0pt] (img3) {\includegraphics[width=\linewidth]{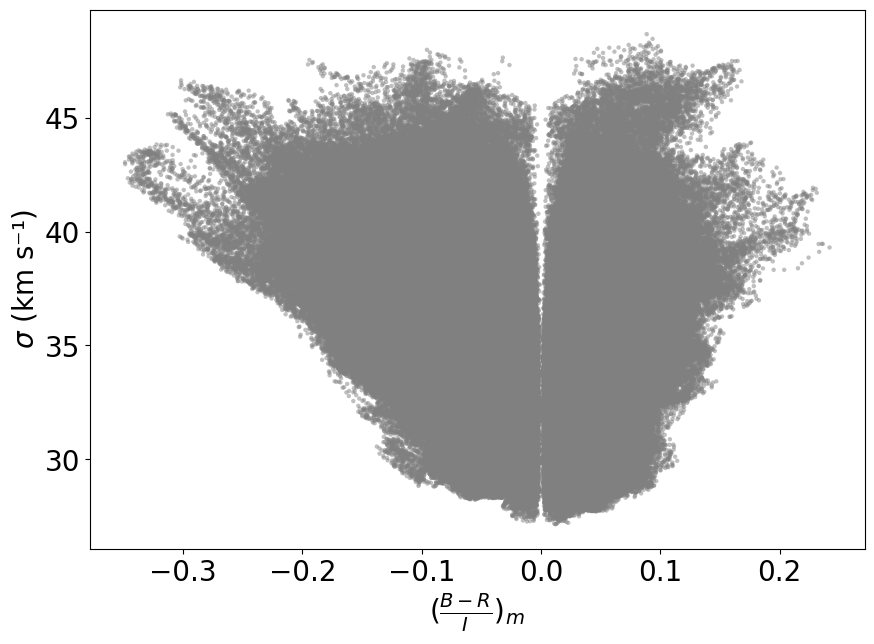}};
            \node[anchor=south west, xshift=-3pt, yshift=10pt] at (img3.north west) {\textbf{(c)}};
        \end{tikzpicture}
    \end{minipage}%
    \hfill
    \begin{minipage}{0.48\textwidth}
        \centering
        \begin{tikzpicture}
            \node[inner sep=0pt] (img4) {\includegraphics[width=\linewidth]{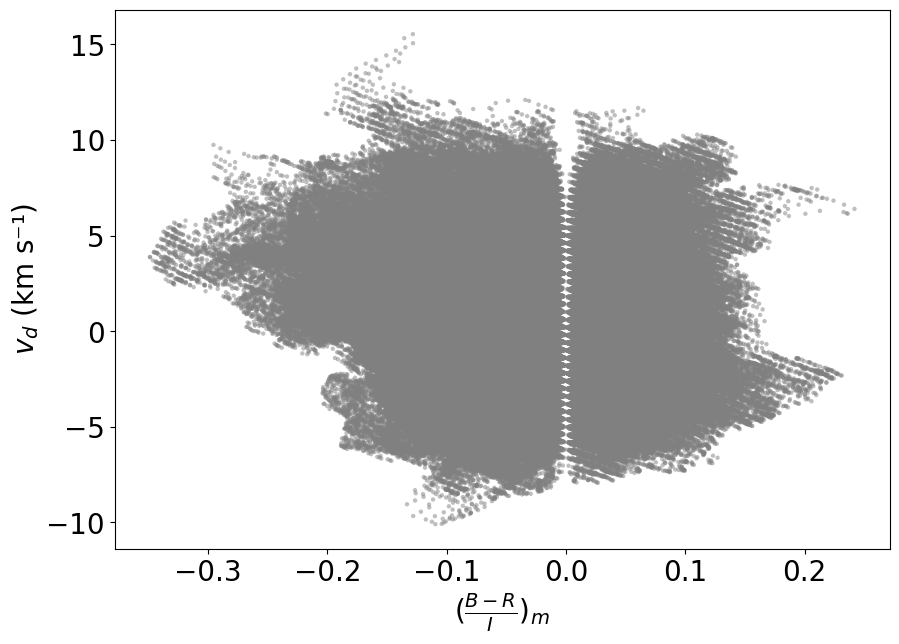}};
            \node[anchor=south west, xshift=-3pt, yshift=10pt] at (img4.north west) {\textbf{(d)}};
        \end{tikzpicture}
    \end{minipage}

    \vspace{0.6cm} 

    \caption{Same as Figure \ref{fig:br_st_l} but for $v_{rms}$ = 15 km s$^{-1}$ and 26 km s$^{-1}$, for emission integrated along different random LOS, where the LOS lies in the direction perpendicular to and along the magnetic field direction.}
    \label{fig:br_st_5}
\end{figure*}

We have also performed a general case analysis combining the different LOS for $v_{rms}$ = 15 km s$^{-1}$ and 26 km s$^{-1}$, the lines of sight are randomly chosen across and along the magnetic field direction. Here too, the magnitude of BR asymmetry is reaching up to 35\% on the redward side while around 25\% on the blueward side. The bimodal distribution is less symmetric, but both asymmetries could be seen. The velocity distribution for maximum asymmetry peaks at around 45-50 km s$^{-1}$. The line widths are in the range 25- 45 km s$^{-1}$, and the Doppler shifts are around $\pm 10$ km s$^{-1}$.

\bibliography{sample631}{}
\bibliographystyle{aasjournal}



\end{document}